\documentclass[12pt]{JHEP}

\def\lsim{\raise0.3ex\hbox{$\;<$\kern-0.75em\raise-1.1ex
\hbox{$\sim\;$}}}
\def\gsim{\raise0.3ex\hbox{$\;>$\kern-0.75em\raise-1.1ex
\hbox{$\sim\;$}}}

\title{
Exploring Neutrino Mixing with Low Energy Superbeams
}

\author{
Hisakazu Minakata \\
Department of Physics, Tokyo Metropolitan University \\
1-1 Minami-Osawa, Hachioji, Tokyo 192-0397, Japan\\
E-mail: \email{minakata@phys.metro-u.ac.jp}
}

\author{Hiroshi Nunokawa\\
Instituto de F\'{\i}sica Gleb Wataghin, Universidade Estadual
de Campinas - UNICAMP,
P.O. Box 6165, 13083-970 Campinas SP Brazil\\
E-mail: \email{nunokawa@ifi.unicamp.br}
}

\abstract{
We explore as clearly as possible the features of neutrino
oscillation which are relevant for measurements of
the CP violating Kobayashi-Maskawa phase $\delta$ and
the sign of $\Delta m^2_{13}$.
We focus on the so called low-energy option and discuss principles
for optimizing experimental parameters to measure these two quantities
simultaneously.
Toward the goal, we first formulate a method for obtaining a bird-eye
view of the phenomenon of neutrino oscillation by introducing a new
powerful tool called the ``CP trajectory diagram in bi-probability space''.
It allows us to represent pictorially the three effects separately
in a single diagram;
effect from genuine CP violation due to the $\sin \delta$ term,
effect from the CP conserving $\cos \delta$ term,
and the fake CP violating effect due to earth matter.
By using the CP trajectory diagram we observe that there is a
two-fold ambiguity in the determination of $\delta$ which is
related with the sign of $\Delta m^2_{13}$.
We then address the question of what are the promising options
for conceptual design of experiments at low energies which looks
for CP violation and at the same time would resolve the
two-fold ambiguity.
We point out that a version with distance of about 700 km,
CERN to Gran Sasso and/or Fermilab to Soudan-2 site, with
a megaton class water Cherenkov detector gives an optimal
design which allows simultaneous determination of $\delta$
and the sign of $\Delta m^2_{13}$ {\it in situ}.
We also point out that there is a possibility that the similar
{\it in situ} measurement of both quantities can be done at
the Phase II of JHF
experiment with much shorter baseline, under the assumption of
nature's kind setting of $\delta$ to the region of
$\sin{\delta} \cdot \Delta m^2_{13} < 0$.
A technique of running at high ($\sim 1$ GeV) and low ($\sim 0.5$ GeV)
beam energies is proposed as a method for better identification
of $\delta$.
}


\begin{document}
\input psfig

\section {Introduction}

Discovery of neutrino oscillation in atmospheric neutrino observation
in the Super-Kamiokande (SK) experiment~\cite {SKatm}
opened up a new window to physics
beyond the standard model of particle physics. Moreover, robust
discrepancy of the measured flux of solar neutrinos to the calculated
one \cite {solar} presents another indication for neutrino masses and
the lepton flavor mixing.
In fact, the first result from the SNO~\cite{SNO} group combined
with the results by the Super-Kamiokande solar neutrino observation
\cite{SKsolar} strongly indicates that solar neutrinos also do oscillate.
The existence of phenomenon of neutrino oscillation is further
strengthened by the result of the K2K experiment, the first
long-baseline experiment with artificial neutrino beam, in particular
by their latest result \cite{K2K}.

The determination of the complete structure of the lepton flavor mixing
matrix, the Maki-Nakagawa-Sakata (MNS) matrix \cite {MNS}, is one of
the most challenging task in particle physics.
While we began to grasp the values of the leptonic mixing parameters
there remain three quantities which are poorly known
or not constrained at all at this moment.
They are $\theta_{13}$, the sign of $\Delta m^2_{13}$,
and $\delta$, the leptonic Kobayashi-Maskawa
angle \cite{KM,early}.
(See e.g. Ref. \cite {NOW2000mina} for a summary of situations
of yet to be determined parameters in the three-flavor mixing
scheme of neutrinos.)
Regarding to $\theta_{13}$, we only know its upper limit,
$\sin^2 2\theta_{13}\lsim 0.1$, from the reactor experiments~\cite{CHOOZ}.
The bound is to be improved, or the value of $\theta_{13}$ itself
could be determined by the next generation long baseline experiments
\cite {JHF,MINOS,OPERA}.
About the sign of $\Delta m^2_{13}$,
while there is a strong indication that the inverted mass
hierarchy (or negative $\Delta m^2_{13}$) is disfavored by
the observed neutrino events coming from supernova SN1987A~\cite {MN01},
no hint is available from laboratory experiments.
With respect to CP phase, $\delta$, we do not have any experimental
clue at all.

In this paper, we intend to explore as clearly as possible the features
of neutrino oscillation which are relevant for experimental
measurements of the latter two quantities,
$\delta$ and the sign of $\Delta m^2_{13}$.
We focus on the so called low-energy option
\cite {NOW2000mina} and discuss principles for optimizing
experimental parameters to measure these two quantities simultaneously.
Toward the goal, we first formulate a method for obtaining a
bird-eye view of the phenomenon of neutrino oscillation at low
neutrino energies, typically, 0.5-2.0 GeV.
We do this by introducing a new powerful tool called the
``CP trajectory diagram in bi-probability space'',
which allows us to represent pictorially the effects of genuine
CP violating phase and the earth matter \cite {MSW} separately
in a single diagram, as explained in detail below.
We then address the question of what are the promising options
for conceptual design of experiments which look for CP violation
among thinkable varying possibilities at low energies.

We have discussed in previous communications \cite{MN00,Nufact00nuno}
the possibility of use of an intense low energy neutrino beam,
which is nowadays referred to as a ``superbeam'' \`a l\`a Richter
\cite {richter}, to measure CP violation.
It is based on the underlying cancellation mechanism of the
leading-order matter effect, the vacuum mimicking mechanism
\cite{mimicking},
which would allow us to measure CP violation in a clean
vacuum-effect-dominated environment for neutrino oscillations.
It gives us a great merit of avoiding the notorious problem of
matter effect contamination \cite{AKS97,MN97,MN98,recent}.

As a concrete example which realizes our basic idea we designed
and examined an experiment which uses a neutrino superbeam
of energy $E \simeq 100$ MeV and a 1 megaton water Cherenkov
detector~\cite{MN00,Nufact00nuno}.
We have ended up with a rather short baseline, $L \sim 30-40$ km,
as optimal distance in such an experiment.
While it served well for our purpose of illuminating the basic idea,
several experimental problems have been raised when it was
taken at face value as an experimental proposal.
(See, however, \cite {cadenas} for a detailed feasibility study
of the very similar idea, $E=250$ MeV, $L \sim 100$ km, and a
40 kton water Cherenkov or liquid scintillator detectors.)

If we simply scale up the energy range to the more realistic one,
$E \sim$ 1 GeV, keeping $E/L$ fixed as in
Ref.~\cite{MN00,Nufact00nuno}, the optimal distance may become
naively $L \sim$ 300-400 km. It is very similar to the experimental
set up recently discussed by the JHF neutrino experimentalists.
In particular, by the JHF neutrino working group, various options
for neutrino beam, wide band (WB) beam, narrow band (NB) beam,
and off-axis (OA) beam with neutrino energies of 500 MeV-a few GeV
have been extensively studied \cite {kobayashi}.
Moreover, the efficiency of removing $\pi^0$ contamination
has been improved tremendously since the old Letter of Intent (LOI)
\cite {JHF} by implementing severe cut by imposing a second ring,
a new technique originally developed by the Super-Kamiokande group
\cite {obayashi}. See their new version of LOI in \cite {JHF2}.

In this paper we try to make a step forward along the line of
thought toward measuring CP violation with use of low energy
conventional superbeam by describing a general strategy of
optimizing beam energy and/or baseline distance.
We will show that the matter effect are comfortably large
even in these mediumly long baseline ($\sim 300$ km) experiments
so that there is a possibility of simultaneous measurement of
$\delta$, the leptonic Kobayashi-Maskawa angle, and the sign of
$\Delta m^2_{atm}$ {\it in situ} in a single experiment.

An alternative strategy based on intense neutrino beam from
a muon storage ring called neutrino factory has been extensively
discussed in the literature \cite {nufact1,nufact2,KS00}.
Our strategy which utilizes low energy conventional superbeam
differs in many ways from neutrino factory, e.g., on beam energy,
detection principle, and most crucially, on whether the value of
$\theta_{13}$ must be known in advance (conventional beam),
or can be measured simultaneously with $\delta$ (neutrino factory).
While many discussions are going on upon ignition by Ref.
\cite {richter} about which strategy is more superior
\cite{nufact.vs.sb},
we strongly feel that both strategies must be fully
developed both in physics and beam technology aspects before
attempting any real comparisons between the two strategies.

The motivation for our consideration of neutrino experiments
with beam energies of $\sim 1$ GeV in this paper is partly
theoretical and partly experimental. Experimentally, the energy
region comes out as a natural compromise of the two requests that
CP violation is large and neutrino beam is intense enough.
Furthermore, it has a great merit of being able to utilize the
results of the recent developments we just mentioned above.
On the theoretical side, we discuss below by using the
``CP trajectory diagram'' a principle of optimizing beam energy
and/or path length to maximizes the detection probability of CP
violation.
We will see that such discussion naturally leads to
several options which utilize the same energy region,
$E \sim 1$ GeV.
Notably, we will uncover the possibility in which an {\it in situ}
simultaneous measurement of CP violating phase $\delta$ and the sign
of $\Delta m^2_{13}$ will be possible for
(i) whole region of $\delta$ ($L \simeq 700$ km), and
(ii) half a region of $\delta$ which fulfills the condition
$\sin{\delta} \cdot \Delta m^2_{13} < 0$ ($L \simeq 300$ km).

In Sec. 2 we point out that an approximate two-fold degeneracy
exists in vacuum neutrino oscillation probability and show that
the degeneracy is partly lifted by the matter effect.
We introduce in Sec. 3 the CP trajectory diagram on bi-probability
space, and thoroughly analyze its properties.
It will be shown that it is a powerful tool for illuminating the
general structure of the oscillation probabilities of
neutrinos and antineutrinos.
We will point out that the two-fold degeneracy which exists
in vacuum neutrino oscillation probability,
after resolved by matter effect, leaves a remnant
ambiguity in the determination of CP violating phase $\delta$.
We then discuss in Sec. 4 a principle of tuning beam energies
to have maximal CP violation as well as to help in resolving 
the two-fold ambiguity.
In Sec. 5 we describe several possible ideas for resolving
the two-fold ambiguity.
Throughout these sections we will reveal that a new strategy
toward simultaneous determination of the CP violating angle
$\delta$ and the sign of $\Delta m^2_{13}$ naturally emerges
from the discussions of problems mentioned above.
In Sec. 6 we describe some concrete examples of experiments which
utilize conventional neutrino superbeams. We estimate the number of
events and include the background rate to obtain a rough idea
for the accuracy of the measurement.
In Sec. 7 we briefly discuss the CP trajectory diagram with experimental
parameters for neutrino factory.
In Sec. 8 we state our conclusions.

\section {Matter effect helps to resolve two-fold ambiguity in vacuum
neutrino oscillation}

We start by describing a role played by matter effect to help resolve
a two-fold ambiguity which would exist in a vacuum-effect dominated
neutrino oscillation experiment to measure CP violating angle $\delta$.
It contrasts to the negative role played by the matter effect as a
contamination in measurement of genuine CP violation, the widely
recognized fact in the literature \cite{AKS97,MN97,MN98,recent}.
We hope that the discussion illuminates the
necessity of complete understanding of the interplay between effects
due to the Kobayashi-Maskawa phase and earth matter.

We use throughout this paper the standard notation of the MNS matrix:
\begin{equation}
U=\left[
\begin{array}{ccc}
c_{12}c_{13} & s_{12}c_{13} &   s_{13}e^{-i\delta}\nonumber\\
-s_{12}c_{23}-c_{12}s_{23}s_{13}e^{i\delta} &
c_{12}c_{23}-s_{12}s_{23}s_{13}e^{i\delta} & s_{23}c_{13}\nonumber\\
s_{12}s_{23}-c_{12}c_{23}s_{13}e^{i\delta} &
-c_{12}s_{23}-s_{12}c_{23}s_{13}e^{i\delta} & c_{23}c_{13}\nonumber\\
\end{array}
\right].
\label{MNSmatrix}
\end{equation}
We define the neutrino mass-squared difference as
$\Delta m^2_{ij} \equiv m^2_{j} - m^2_{i}$ where
$\Delta m^2_{12}$ is assumed to be smallest,
relevant for oscillation solutions to
the solar neutrino problem.
We note in passing that in the MSW mechanism,
under the assumption that $\Delta m^2_{23} \sim \Delta m^2_{13}$
is much larger than $\Delta m^2_{12}$,
latter can be made always positive as far as $\theta_{12}$ is taken in
its full range [0, $\pi/2$]\cite {FLMP}.

The ambiguity arises due to the fact that $\Delta m^2$ scale implied
by solar neutrino solution may be small such that
\begin{equation}
\frac {\Delta m^2_{12} L}{2 E} = 2.54 \times 10^{-2}
\left(\frac{\Delta m_{12}^2}{10^{-5}\ \mathrm{eV}^2}\right)
\left(\frac{L}{1000\ \mathrm{km}}\right)
\left(\frac{E}{1\ \mathrm{GeV}}\right)^{-1}
\label{smparam}
\end{equation}
is sufficiently small compared to unity for most of the possible
experimental parameters.
To first order in the parameter the neutrino oscillation
probabilities of
$\nu_{\mu} \rightarrow \nu_e$ and
$\bar{\nu}_{\mu} \rightarrow \bar{\nu}_e$
in vacuum are given by
\begin{eqnarray}
P_{vac \pm}[\nu_{\mu}(\bar{\nu}_{\mu})
\rightarrow \nu_{\rm e}(\bar{\nu}_e)]
&=&
\sin^2{2\theta_{13}} s^2_{23}
\sin^2 \left(\frac{\Delta m^2_{13} L}{4 E}\right)
\nonumber \\
&-& \frac {1}{2}
s^2_{12}\sin^2{2\theta_{13}}s^2_{23}
\left(\frac{\Delta m^2_{12} L}{2 E}\right)
\sin \left(\frac{\Delta m^2_{13} L}{2 E}\right)
\nonumber \\
&+&
2J_{r} \cos{\delta}
\left(\frac{\Delta m^2_{12} L}{2 E} \right)
\sin \left(\frac{\Delta m^2_{13} L}{2 E}\right)
\nonumber \\
&\mp&
4J_{r}\sin{\delta}
\left(\frac{\Delta m^2_{12} L}{2 E}\right)
\sin^2 \left(\frac{\Delta m^2_{13} L}{4 E}\right),
\label{Pvac}
\end{eqnarray}
where $-$ and $+$ sign in front of the 4th term 
refer to the neutrino and the antineutrino channels, respectively, 
and $J_{r}=c_{12}s_{12}c^2_{13}s_{13}c_{23}s_{23}$ denotes
the reduced Jarlskog factor.
One can readily observe that the oscillation probabilities,
except for the second term, are invariant under
simultaneous transformation
\begin{eqnarray}
& &\delta \rightarrow \pi - \delta
\hskip 0.5 cm (\mbox{mod.} 2 \pi),
\nonumber \\
& &\Delta m^2_{13} \rightarrow - \Delta m^2_{13}.
\label{flipsym}
\end{eqnarray}
Under the transformation (\ref{flipsym}), flipping sign of
the $\cos{\delta}$ term by the first transformation is
canceled by the second, whereas
the $\sin{\delta}$ term is manifestly invariant. It implies that
the probability is approximately degenerate for two values of $\delta$,
unless one know {\it a priori} the sign of $\Delta m^2_{13}$,
and hence there is a two-fold ambiguity in determination of $\delta$.

We note that the invariance holds in a very good approximation.
It is because the second term is small in
magnitude because of the suppression factors including
$\left(\frac{\Delta m^2_{12} L}{2 E}\right)$ and the fact that
$\sin^2{2 \theta_{13}}$ is smaller than 0.1 
in order to satisfy the CHOOZ result~\cite {CHOOZ}. 
Also the higher order terms
of $\left(\frac{\Delta m^2_{12} L}{2 E}\right)$ is sufficiently
small for most of the experimental settings.
It may be worthwhile, however, to note that the degeneracy is
accidental and approximate in nature, and represents neither inherent
nor exact properties of vacuum neutrino oscillations.

Of course, the ambiguity does not exist if we know in advance
the sign of $\Delta m^2_{13}$.
However, it is very unlikely that the question of the normal vs.
inverted mass hierarchies of neutrinos will be answered
in a convincing way in the near future.
It is because (as it is believed)
the determination of the sign of $\Delta m^2_{13}$ requires
measurement of interference between the CP and the matter
effects, which necessitates a sufficiently long baseline.
It is not an easy experiment to carry out because it requires
either intense neutrino beam or supermassive detectors, or
plausibly both.

We now point out that the two-fold degeneracy in vacuum oscillation
probability is lifted by the matter effect represented by
the index of refraction $a(x) = \sqrt {2} G_F N_e (x)$
where $N_e (x)$ is the electron number density in the earth.
When we include the matter effect there arise,
in leading order of $aL$, the following additional terms $P_{matt}$
in the oscillation probability computed under the adiabatic
approximation \cite {AKS97}:
\begin{eqnarray}
P_{matt \pm}[\nu_{\mu}(\bar{\nu}_{\mu})
\rightarrow \nu_{\rm e}(\bar{\nu}_e)]
&=&
\pm \cos{2\theta_{13}}
\sin^2{2\theta_{13}} s^2_{23}
\left(\frac {2 Ea(x)}{\Delta m^2_{13}}\right)
\sin^2 {\left(\frac{\Delta m^2_{13} L}{4 E}\right)}
\nonumber \\
&\mp&
\frac{a(x)L}{4}\sin^2{2\theta_{13}}\cos{2\theta_{13}} s^2_{23}
\sin \left(\frac{\Delta m^2_{13} L}{2 E}\right),
\label{Pmatt}
\end{eqnarray}
In (\ref{Pmatt}), $a(x)= \sqrt 2 G_F N_e(x)$
where $G_F$ is the Fermi constant and $N_e(x)$ denotes the
electron number density at $x$ in the earth,
and $+(-)$ in the 1st term and $-(+)$ in the 2nd term 
refer to the neutrino (antineutrino) channel. 

It is easy to observe that the degeneracy is lifted because
$P_{matt}$ is not invariant under flipping sign of $\Delta m^2_{13}$,
a well known fact. What is perhaps not so well known is that
the lifting of the degeneracy does not completely resolve
the two-fold ambiguity in determination of $\delta$.
We introduce in the next section a powerful tool called the
``CP trajectory diagram'' and demonstrate that a remnant ambiguity
exists in the determination of CP violating phase.
Since the ambiguity is related with the sign of $\Delta m^2_{13}$,
we are naturally invited to the problem of simultaneous
determination of $\delta$ and $\Delta m^2_{13}$, which we will pursue
in Sec. 6.

\section {CP trajectory diagram in bi-probability space}

To illuminate global features of neutrino oscillations relevant
for low energy experiments, we introduce the CP trajectory diagram
in bi-probability space spanned by $P(\nu)$ and $P(\bar{\nu})$
\cite {mina.nufact01}.
Unless otherwise stated we simply denote
$P(\nu_{\mu} \rightarrow \nu_{e})$ and
$P(\bar{\nu}_{\mu} \rightarrow \bar{\nu}_{e})$
as $P(\nu)$ and $P(\bar{\nu})$ in this paper.
We show in this section that the diagram is a useful tool for
our purpose because it can display pictorially the three effects
separately in a single diagram; genuine effect of CP violating
phase $\delta$ coming from $\sin\delta$ term,
CP conserving effect due to $\cos\delta$ term,
and the matter effect.

Suppose that we compute the oscillation probability $P(\nu)$ and
$P(\bar{\nu})$ with a given set of oscillation and experimental
parameters. Then, we draw a dot on two-dimensional plane spanned
by $P(\nu)$ and $P(\bar{\nu})$.
When $\delta$ is varied we have a set of dots which forms a closed
trajectory, closed because the probability must be a periodic
function of $\delta$, a phase variable.

In Fig. 1 plotted is the contours of oscillation probabilities
$P(\nu)$ and $P(\bar{\nu})$ which are drawn by varying
the CP violating phase, $\delta$, from 0 to $2\pi$.
We note that in this work, while we refer to analytic expressions
for the explanations, 
all the results shown in our plots as well as in tables were 
based on the computations obtained by numerically solving 
the three flavor neutrino evolution equation assuming 
constant matter density. 
The solid and the dashed lines are for positive and negative
$\Delta m^2_{13}$, respectively.
The dotted and the dash-dotted curves are the cases in vacuum in which
the matter effect is switched off in the corresponding cases of the
solid and the dashed lines, respectively.
In fact, the abscissa and the ordinate are not quite the oscillation
probabilities but are the ones averaged over an
appropriate energy distribution of (anti-) neutrinos.
It is to avoid accidental zeros, and at the same time is also meant
to mimic the average over the energy dependent flux times cross
sections, the procedure we will actually execute in Sec. 6.
The neutrino energy distribution is taken to be a Gaussian shape
with central value of (a) 0.5 GeV, (b) 1.0 GeV,  (c) 1.5 GeV,
and (d) 2.0 GeV, respectively.
The widths of Gaussian distribution is taken to be 20 \% of
the peak energies, e.g., 100 MeV for $E=500$ MeV in Fig.1a.
The baseline length is taken to be 295 km, JHF-Kamioka distance,
and the values of other parameters are typical ones for the large
angle MSW solution to the solar neutrino problem and
$\sin^2{2 \theta_{13}}=0.05$, as given in the caption of Fig.1.
Notice that unless the large angle MSW solution is the case
the measurement of CP violation would be very difficult.

\FIGURE[!ht]{
\centerline{\psfig{file=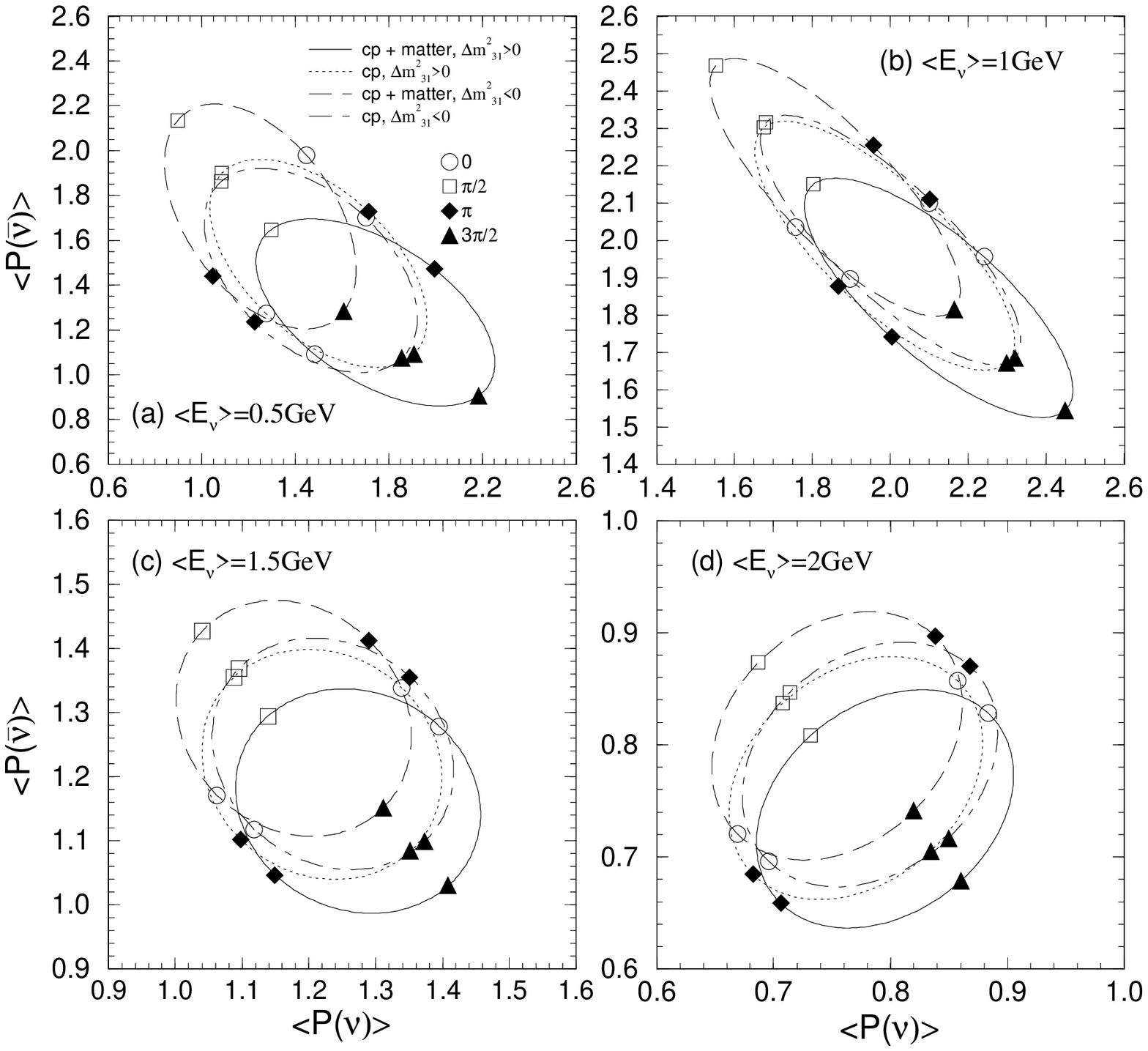,height=15.5cm,width=16.5cm}}
\vspace{-1.3cm}
\caption{
CP trajectory in the bi-probability (given in \%) plane
for the baseline $L=295$ km.
As indicated in the figures, the solid and the dashed lines are for
$\Delta m^2_{13} > 0$ and $\Delta m^2_{13} < 0$ cases, respectively,
and the dotted and the dash-dotted lines correspond to the same
signs of $\Delta m^2_{13}$ as above but with matter effect switched off.
The mixing parameters are fixed as
$\Delta m^2_{13}
 = \pm 3 \times 10^{-3}$ eV$^2$,
$\sin^22\theta_{23} = 1.0$,
$\Delta m^2_{12} = 5\times 10^{-5}$ eV$^2$,
$\sin^22\theta_{12} = 0.8$,
$\sin^22\theta_{13} = 0.05$.
We take $\rho Y_e = 1.4$ g/cm$^3$ where $\rho$ is
the matter density and $Y_e$ is the electron fraction.
}
\label{Fig1}
}

As one might have suspected by looking in Fig.1 the CP
trajectory is elliptic. It is easy to prove it in the vacuum
case and it can be also shown that it is the case in a good
approximation for oscillations in matter. The proof of the
statements will be given in Appendix.
In particular, it is shown that the major (or minor) axis
is always at 45 degree in the vacuum case.

What does CP trajectory diagram actually represent?
To answer the question, we start with the vacuum case and
first concentrate on Fig.1a-c.
In these cases, the lengths of the major and the minor axes
are measures for the coefficients of $\sin{\delta}$ and
$\cos{\delta}$, respectively, in the oscillation probability.
(The same statement holds for Fig.1d if the major and the minor
axes are interchanged.)
One can readily understand this statement by looking at
Eq. (\ref{ellip.vac}) in Appendix.

The two trajectories of positive and negative $\Delta m^2_{13}$
which are represented by dotted and the dash-dotted lines,
respectively, are almost degenerate. One notices that the approximate
degeneracy is between $\delta$ and $\pi - \delta$ (mod. 2 $\pi$)
cases as we anticipated in discussions in the previous section.
The two trajectories slightly split mainly due to the second
term in Eq.~(\ref{Pvac}) and possibly by higher order terms of
$\left(\frac{\Delta m^2_{12} L}{2 E}\right)$
which are not taken into account in Eq.~(\ref {Pvac}).

In matter, the CP trajectories of neutrinos and antineutrinos
split; the former moves to downward-right ($\Delta m^2_{13} > 0$) and
the latter to upward-left ($\Delta m^2_{13} < 0$).
In fact, one can explicitly demonstrate that the matter effect is
the cause of the departure of the two trajectories by artificially
increasing the matter potential $a$.
In Fig. 2, presented are the
results of the same computations as in Fig. 1 but with a factor
of 2 (artificially) larger matter effect. One observes that the
neutrino and the antineutrino trajectories became more separated
along the direction of major (minor) axis on bi-probability plane
in Fig.2a-c (Fig.2d).
Therefore, the degree of nonoverlapping of neutrino and
antineutrino trajectories gives almost purely the measure for
the matter effect.
%

\FIGURE[!ht]{
\centerline{
\psfig{file=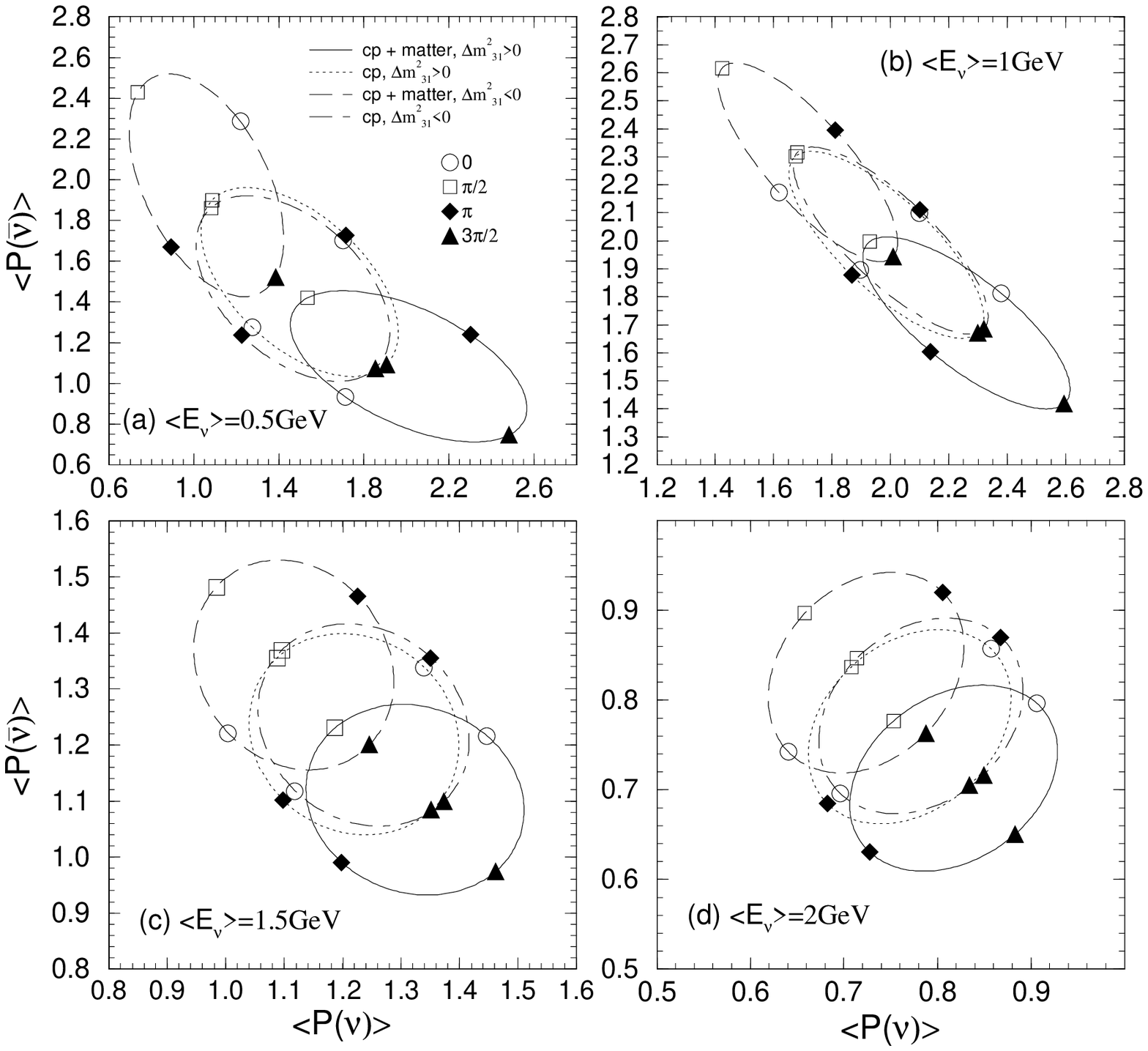,height=15.5cm,width=16.5cm}}
\vspace{-1.3cm}
\caption{
The same as in Fig.\ref{Fig2} but for $\rho Y_e = 2.8$ g/cm$^3$.
}
\label{Fig2}
}

It is important to notice, by comparing the cases in vacuum
and in matter in each figure, that lengths of the major and the
minor axes, which measure the effect of $\sin{\delta}$ term,
the genuine CP violation, and the $\cos{\delta}$ term, respectively,
barely change by the matter effect. It is nothing but the feature
which we expect from the perturbative formula
in Eqs.~(\ref{Pvac}) and (\ref{Pmatt}).

Thus, the matter effect lifts the two-fold degeneracy of $\delta$
and $\pi - \delta$ cases discussed in Sec. 2, as shown in Fig. 1.
Notice, however, that there still remains two crossing points
of the two trajectories corresponding to positive and negative
$\Delta m^2_{13}$ as indicated in Fig. 1.
It means that if we are unlucky so that the true value of
$\delta$ is close to the crossing point, then we will still
have two-fold degeneracy in determination of $\delta$.
Since we expect only a modest statistics even with a huge detector
in experiments for measuring $\delta$, the region suffered by the
ambiguity problem may not be so small,
unless nature was so kind to tune her parameters to produce
maximal CP violating effects. We will discuss in the next section
how to resolve the remaining two-fold ambiguity.

We emphasize that it is one of the nicest features of the
CP trajectory diagram that the CP violating as well as
conserving effects due to the Kobayashi-Maskawa phase, and
the earth matter effect are pictorially displayed separately
in a single diagram.


The readers may worry about the apparently intricate features of
the CP trajectory diagram indicated in Figs. 1 and 2. They include:

\vskip 0.3cm
\noindent
(F1) the feature that the ``chirality'' of the trajectory depends
upon the sign of $\Delta m^2_{13}$. Here, what we mean by
``chirality'' is how a trajectory winds, clockwise or
counter-clockwise, as $\delta$ varies from 0 to 2$\pi$.

\vskip 0.3cm
\noindent
(F2) the dependence on the neutrino energy; between energies
of 0.5 GeV and 1 GeV the positions of $\delta = 0$ and $\delta = \pi$
are exchanged completely.
At the same time, the ``chirality'' of the trajectory also flips
from 0.5 GeV to 1 GeV.

\vskip 0.3cm

The explanation of (F1) and (F2) are in fact very simple.

\vskip 0.3cm

\noindent
(A1) Since the $\delta$-development of the trajectory is uniform,
it suffices to discuss the behavior of the trajectory at around
$\delta = 0$. When $\delta$ increases from zero, the $\sin{\delta}$
term in the oscillation probability (\ref{Pvac}) decreases (increases)
in neutrino (antineutrino) channel.
This means that the movement of the trajectory is toward
upward-left direction for both $\Delta m^2_{13} > 0$ and
$\Delta m^2_{13} < 0$ cases.

The trajectory winds in opposite way for positive and negative
$\Delta m^2_{13}$ because the point $\delta = 0$ is located
at near (far) side from the origin for $\Delta m^2_{13} > 0$
($\Delta m^2_{13} < 0$) case,
due to the $\cos{\delta}$ term in the probability (\ref{Pvac}),
at $E = 0.5$ GeV as in Fig.1a.
It means that the trajectory winds clockwise for
$\Delta m^2_{13} > 0$ and winds counter-clockwise for
$\Delta m^2_{13} < 0$.

\vskip 0.3cm

\noindent
(A2) When neutrino energy increases from $E = 0.5$ GeV to 1 GeV,
$\sin \left(\frac{\Delta m^2_{13} L}{2 E}\right)$ changes sign
from negative to positive at $L = 300$ km.
Then, the $\cos {\delta}$ term flips the sign and the
$\delta = 0$ point jumps to the far (near) side of the trajectory
in $\Delta m^2_{13} > 0$ ($\Delta m^2_{13} < 0$) case.
Then, the CP trajectories winds to the opposite directions with
those at $E = 0.5$ GeV.

\section {Principle of choosing beam energies for long-baseline
neutrino oscillation experiments}

The intriguing features of the CP trajectory diagram mentioned
above implies a new principle for determining neutrino beam
energies for given oscillation parameters and the baseline length.

To avoid a confusion in notations that might occur when a major
and a minor axes switch with each other depending upon the
parameters we use in this section the terms the ``radial thickness''
and the ``polar thickness'' of contours. In Fig.1a-c (Fig.1d),
the polar thickness denotes the length of major (minor) axis,
while the radial thickness implies that of minor (major) axis.

As explained in the last section, the radial and the polar thickness
of the CP trajectory diagram are the measure for $\cos{\delta}$
and $\sin{\delta}$ terms in the oscillation probability, respectively.
Now we try to maximize these thickness by tuning experimental
parameters for a given set of mixing parameters. It is of course
important to have a large $\sin{\delta}$ term because it is
the signal for genuine CP violation. On the other hand,
to maximize the radial thickness is to help in resolving the problem of
two-fold ambiguity which was discussed in the last two sections.
We believe that the latter is important in view of uncertainties due to
statistical and systematic errors which would inevitably exist in
any experiments.
If we make a choice of, for example, $E=750$ MeV for
$L=300$ km, then the contour shrinks to be approximately
one-dimensional (see Fig.6b) and there is no way to resolve
the two-fold ambiguity~\cite{konaka.strategy}.

We note that in Eq. (\ref{Pvac}) the coefficients of the
$\cos{\delta}$ and the $\sin{\delta}$ terms are proportional
to $x \sin{x}$ and $x \sin^2({\frac{x}{2}})$, respectively,
where $x = \frac{\Delta m^2_{13}L}{2E}$.
Therefore, a maximum of polar thickness, i.e.,
a maximum of absolute value of $\sin{\delta}$ term,
occurs at half value of $E/L$ of the corresponding maximum
of radial thickness, coefficient of the $\cos{\delta}$ term.
Namely, maximization of these terms implies:
\begin{eqnarray}
\left(\frac{E}{\mbox{1 GeV}}\right)_{\cos{\delta}}
 &=& 1.13, 0.47, 0.29
\left(\frac{L}{300\ \mathrm{km}}\right)
\left(\frac{\Delta m_{13}^2}{3 \times 10^{-3}\ \mathrm{eV}^2}\right), \\
\left(\frac{E}{\mbox{1 GeV}}\right)_{\sin{\delta}}
 &=& 0.62, 0.24, 0.14
\left(\frac{L}{300\ \mathrm{km}}\right)
\left(\frac{\Delta m_{13}^2}{3 \times 10^{-3}\ \mathrm{eV}^2}\right)
\end{eqnarray}
Thus, maxima of polar thickness occurs at relatively low energies,
$(E)_{\sin{\delta}} = 620$ MeV at $L=295$ km and
$(E)_{\sin{\delta}} = 1.5$ GeV at $L=730$ km,
in agreement with the conventional wisdom that CP violation
effects are maximal at low energies. We, however, also emphasize
that extremization of $\cos{\delta}$ term requires about twice larger
values of beam energy for a fixed baseline distance.
It is one of the reasons why we were led to the examination of a
little bit of higher energies compared with that in Ref.
\cite {MN00}, $E =$ 0.5-2.0 GeV region, in this paper.
The energies chosen in Fig.1 turned out to be in the ``right range''
in compromising the requirements of maximizing the polar and the
radial thicknesses of the CP trajectories.

The readers might wonder the possibility that large errors due to
statistical and systematic uncertainties completely invalidate
our principle of optimization of beam energies.
We will demonstrate in Sec. 6 that it does not occur at least for
certain range of reasonable oscillation and experimental
parameters.

\section {Resolving two-fold ambiguity in determination of $\delta$}

We have shown in the previous sections
that there exists a two-fold ambiguity in determination of CP
violating phase $\delta$ due to our ignorance of the sign of
$\Delta m^2_{13}$.
We discuss in this section the problem of how to resolve the
ambiguity.

\subsection {Chance for simultaneous measurement
of $\delta$ and the sign of $\Delta m^2_{13}$ {\it in situ}}

Our foregoing analyses of the two-fold ambiguity have revealed
the intriguing possibility that, if we are lucky, an {\it in situ}
simultaneous determination of $\delta$ and the sign of
$\Delta m^2_{13}$ may be possible even at relatively short
baseline as $L = 300$ km.
Namely, the earth matter effect is comfortably large to split
the two trajectories corresponding to positive and negative
$\Delta m^2_{13}$ such that the both quantities can be measured
simultaneously in a certain range of $\delta$.
Namely, if the angle $\delta$ is in the third or the fourth quadrants
for positive $\Delta m^2_{13}$ (the normal mass hierarchy),
or if $\delta$ is in the first or the second quadrants
for negative $\Delta m^2_{13}$ (the inverted mass hierarchy),
then measurement of $P(\nu_{\mu} \rightarrow \nu_{e})$ and
$P(\bar{\nu}_{\mu} \rightarrow \bar{\nu}_{e})$
in neutrino and antineutrino experiments can determine both
quantities simultaneously.

The statement just made above is a conservative one and
the range of lucky determination of $\delta$ without ambiguity
may extend to wider region, as one notices in Fig. 1.
However, how wide is the region depends upon the mixing parameters
as well as experimental uncertainties. We describe an attempt
toward quantifying it in Sec. 6 by computing numbers of events.

Our above observation sharply contrasts with the conventional belief
that very long baseline as $\gsim 1,000$ km is required for
the determination of the sign of $\Delta m^2_{13}$, and open the door
to a simultaneous measurement of $\delta$ and the sign of
$\Delta m^2_{13}$ {\it in situ} at relatively short 
($\sim 300$ km or so) baseline.

We will further pursue the possibility of simultaneous measurement
of $\delta$ and the sign of $\Delta m^2_{13}$ in a single experiment
which is valid for full range of $\delta$ by considering longer
baseline in Sec. 6.

\subsection {Two-detector method}

However, nature may not be so kind. Namely, if the true value of
$\delta$ is within the experimental uncertainties to one of the
two crossing points of the two trajectories,
we have a two-fold ambiguity in the determination of $\delta$.
In the worst case in which $\delta$ is really close to the
crossing points, one cannot resolve the two-fold ambiguity no
matter how accurate were the measurement.

\FIGURE[!ht]{
\hskip -1cm
\psfig{file=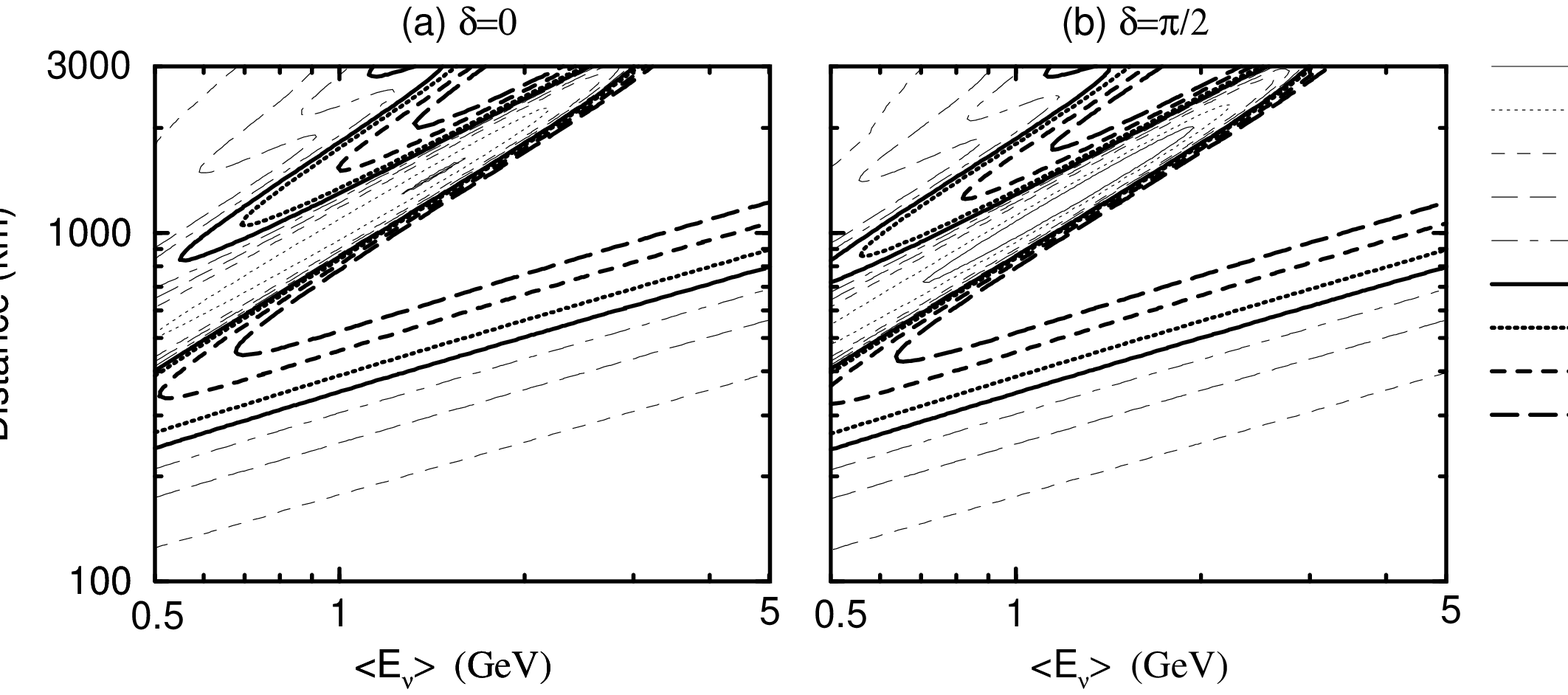,height=17.cm,width=14.cm}
\vspace{-3.0cm}
\caption{
Asymmetry of the probability ratio defined in Eq. (\ref{asymmetry})
in the text for (a) $\delta = 0$ and  (b) $\delta = \pi/2$.
The mixing parameters are chosen as the same with those of Fig. 1.
}
\label{Fig3}
}

Then, the question is how to resolve the two-fold ambiguity.
In this subsection, we discuss the two-detector method in order to
resolve the two-fold ambiguity in determination of the CP violating
angle $\delta$, in case it remains in single-detector experiments.
The multiple detector method has been proposed by various authors
for different physical motivations \cite {2detector}.

We first illuminate that the two-detector option is naturally
motivated by the nature of the phenomenon of neutrino oscillation
itself.
Let us first define the ratio
of the suitably energy averaged appearance
probabilities for neutrino and anti-neutrino, $R(P)$, as follows,
\begin{equation}
R(P) \equiv \frac{\langle P(\nu_\mu \to \nu_e) \rangle}
{\langle P(\bar{\nu}_\mu \to \bar{\nu}_e) \rangle}.
\end{equation}
We note that if there is no CP violating phase and
matter effect, $R(P)$ does neither depend on the baseline
nor on the average neutrino energy as long as energy distributions
of neutrino and anti-neutrino are the same.
Some dependence of $R(P)$ on the baseline can indicate either
genuine CP violating effect or matter effect, or both.
However, by simply looking at $R(P)$, it may be difficult
to separate these two effects.

In order to see for which energy and baseline
the matter effect could be important, independent
of the magnitude of CP phase,
let us define the asymmetry of the ratio
$R(P)$ as follows,
\begin{equation}
A(R) \equiv 2 \times
\frac{R(P;\Delta m_{13}^2>0)-R(P;\Delta m_{13}^2<0)}
{R(P;\Delta m_{13}^2>0)+R(P;\Delta m_{13}^2<0)}.
\label{asymmetry}
\end{equation}
A large value of the asymmetry implies that the matter effect is
enhanced relative to the vacuum effect.
In fact, if there is no matter effect,
from the expression of vacuum probability (\ref{Pvac}),
this quantity is expected to be small for any values of $\delta$.

\FIGURE[!ht]{
\centerline{\protect\hbox{
\psfig{file=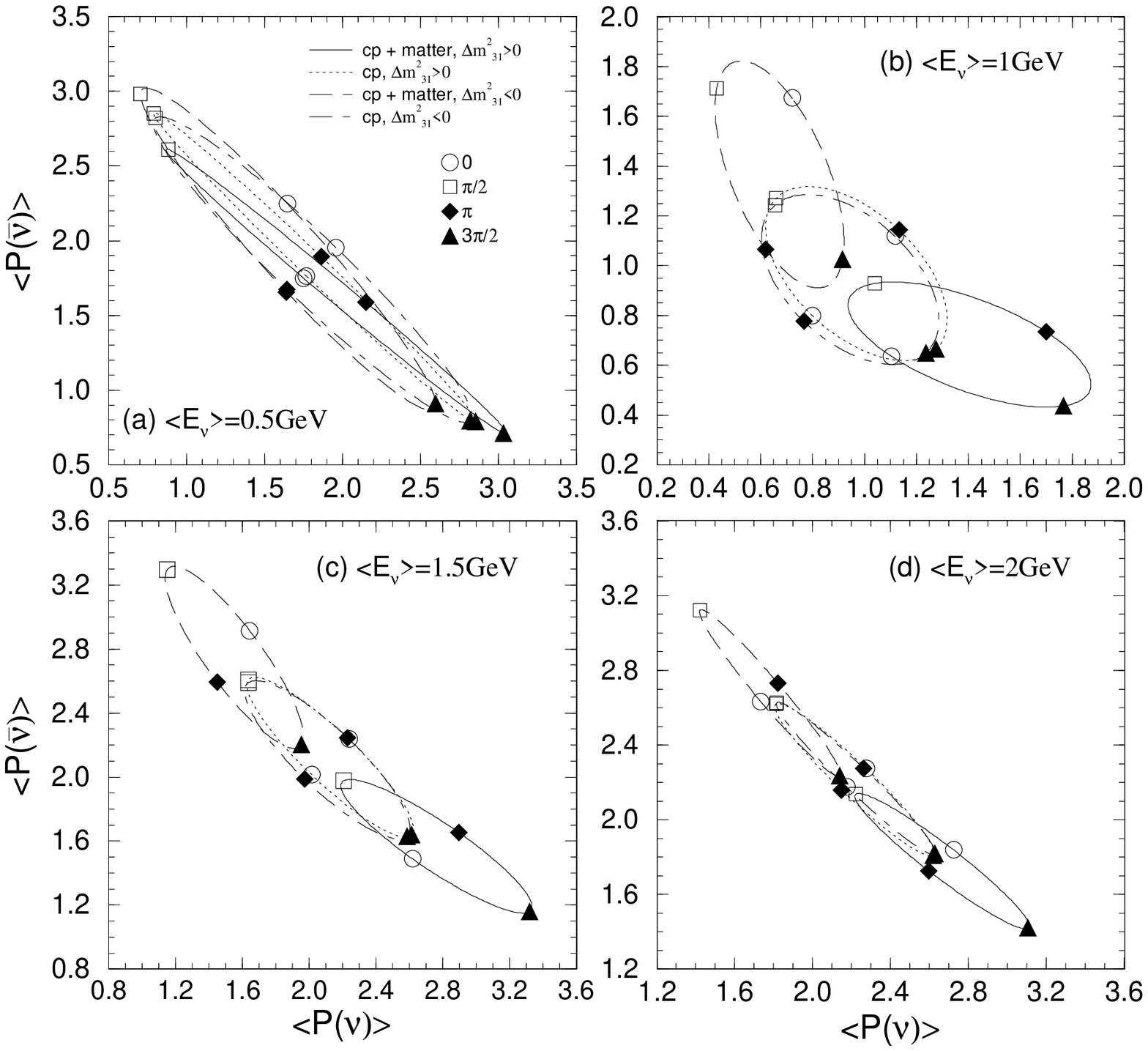,height=15.5cm,width=16.5cm}
}}
\vspace{-1.3cm}
\caption{
The same as in Fig.\ref{Fig2} but for $L = 700$ km.
}
\label{Fig4}
}

In Fig. 3 plotted is the asymmetry $A(R)$ for two typical values
of CP violating angle, $\delta = 0$ and $\delta = \frac {\pi}{2}$.
The mixing parameters are chosen as the same with those of Fig. 1.
There are several characteristic features in the figures.
First of all, as expected, the asymmetry depends very
weakly on $\delta$.
The asymmetry tends to be positive, indicating that
$\Delta m_{13}^2 > 0$ case is enhanced compared to
$\Delta m_{13}^2 < 0$ case.
There are two distinct regions in which the asymmetry $A(R)$ is
large and positive.
At $E \sim 1$ GeV, they are at $L \sim 600 - 700$ km, and at
$1000 - 1500$ km.


\FIGURE[!ht]{
\centerline{\protect\hbox{
\psfig{file=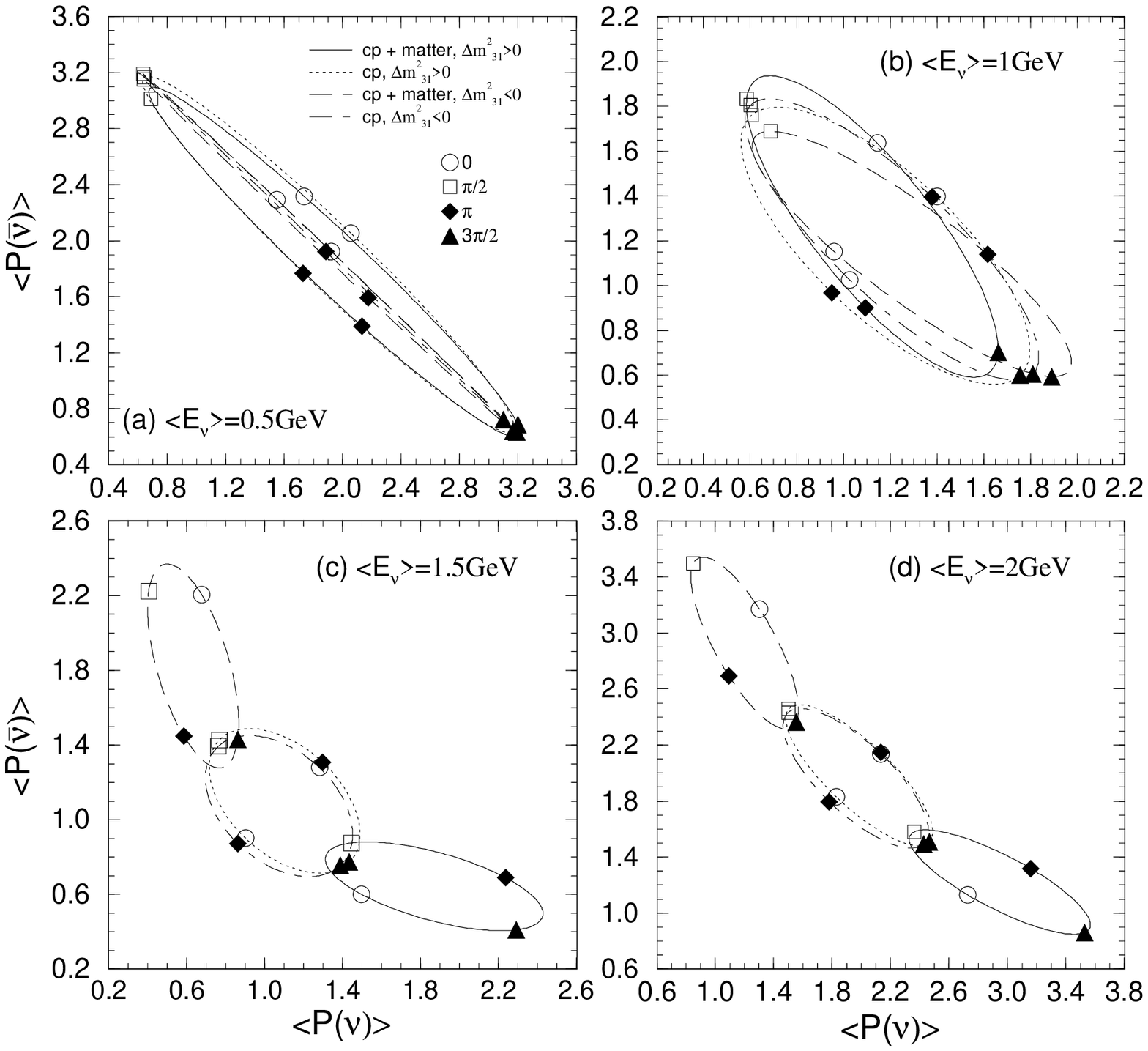,height=15.5cm,width=16.5cm}
}}
\vspace{-1.3cm}
\caption{
The same as in Fig.\ref{Fig2} but for $L = 1000$ km.
}
\label{Fig5}
}

We have repeated the similar computation to obtain the CP
trajectory diagram for the path lengths $L=700$ and 1000 km,
and energies $E=0.5$, 1.0, 1.5, and 2 GeV. The results are
presented in Fig. 4 and 5.
By looking into Figs. 1, 4 and 5,
we immediately observe several distinct features:

\vskip 0.3cm
\noindent
(1) There are cases with longer baselines in which the neutrino
and the antineutrino trajectory do not intersect;
$E = 1.0, 1.5, 2.0$ GeV at $L=700$ km, and
$E = 1.5, 2.0$ GeV at $L=1000$ km.
We note that this is consistent with what we presented
in Fig. 3 where we can see that these values of experimental
parameter fall in the region of strong asymmetry defined in
Eq. (\ref{asymmetry}).
It resolves the two-fold degeneracy we talked about.
Of course, by focusing these experimental parameters, one
can determine $\delta$ and the sign of $\Delta m^2_{13}$
simultaneously, provided that statistical and systematic
uncertainties are small enough.
It raises an important possibility that simultaneous
determination of $\delta$ and the sign of $\Delta m^2_{13}$
can be done in full range of $\delta$ with optimal distance of
$L \simeq 700$ km.
We will further discuss the possibility including experimental
uncertainties in the next section.

\vskip 0.3cm
\noindent
(2) The path-length dependence of the trajectory diagram is
not always smooth. For example, the feature displayed in
$E = 1$ GeV at $L=1000$ km case does not fall into a smooth
extrapolation of the behavior of $L=295$ and 700 km at the
same energy.

\vskip 0.3cm
\noindent
(3) Some curious behavior is observed in the cases,
$E = 0.5$ GeV at $L=700$ and 1000 km, where the matter probabilities
have similar behavior with those of vacuum oscillation,
which seems to be also consistent with the results in Fig. 3.
Most probably, it is a new phenomena, not a remnant of the vacuum
mimicking phenomenon, whose interpretation is not known at the moment.

\vskip 0.3cm

A tentative conclusion before examining the numbers of events
is that by using the second detector placed at distances of 700 km
or 1000 km, the two-fold ambiguity will be resolved if it remains
in shorter baseline JHF-type experiments.

\section {Long-baseline neutrino oscillation experiments
with superbeams; some concrete examples}

In discussions in the foregoing sections we have concentrated
on illuminating global structure of the neutrino
oscillations in the region relevant for low energy experiments,
and did not pay enough attention to the statistical and the
systematic uncertainties.
Of course, they are of key importance in judging
what is the optimal design of the experiments. On the other
hand, it is not quite possible to determine at this moment
which is the optimal design because we do not know the value
of key parameters, most importantly $\theta_{13}$.
Therefore, our discussion in this paper is inevitably restricted
to the one that may be called, at best, case studies.

In this section we first try to estimate the numbers of events
including signal and the background by taking the same typical
mixing parameters as the one we used before, which correspond
to the LMA solution of the solar neutrino problem.
We do this by taking the concrete examples of
upgraded long-baseline neutrino oscillation experiment which
utilizes a conventional neutrino superbeam. For definiteness,
we assume a 4 MW proton beam which is planned to be constructed
in the Phase II of the JHF project, and consider 1 Mton
Super-Kamiokande type water Cherenkov detector whose
fiducial volume is assumed to be 0.9 Mton.
Such detector is already discussed by the Super-Kamiokande
group who mainly motivated by an extensive search for
proton decay under the name of Hyper-Kamiokande
project \cite {shiozawa}.
The goal of our study in the present section is not to develop the
experimental proposal which is ready to submit, but to obtain
a feeling on what are the promising possibilities which deserve
further detailed studies, possibly by joint collaboration by
theorists and experimentalists.

We discuss three options,

\noindent
(i) Single detector at $L=300$ km.

\noindent
(ii) Single detector at $L=700/1000$ km.

\noindent
(iii) Two detectors at $L=300$ and $700/1000$ km.

\noindent

In this work, we will use two types of different neutrino beams
with quasi-monochromatic energy spectrum
calculated by the JHF-SK Neutrino Working Group \cite {kobayashi},
which are (i) narrow band (NB) beam and (ii) off axis (OA) beam.
NB beam is made by pions with particular choice of
momentum selected by dipole magnet placed between
two horns in the wide band beam configuration.
We use in our analysis two different options of NB beam
characterized by pion momenta of 2 and 3 GeV, 
which peaks at $\sim$ 1 GeV and 1.4 GeV, respectively.
OA beam is an another way of making quasi-monochromatic neutrino
beam proposed in Beavis et al. in \cite {2detector}.
It is obtained by slightly (a few degrees) displaced the
direction of axis of wide band beam from the far detector direction.
For this type of beam, we use the ones obtained with
off axis of 3 and 2 degrees, which peaks at 
$\sim$ 0.5 GeV and 0.8 GeV, respectively.
While we adopt as the basic parameters the ones calculated
for the JHF neutrino experiment, we do hope that the results of our
calculation are illuminative enough for future projects on
other continents as well.

\subsection {Method for calculation of number of events}

Before proceeding to physics discussions we have to explain
first how we calculate expected numbers of events.
In our computation, we take into account both signal and background,
which are calculated by the way as explained below. 

Signal consists of the contributions from charged current (CC)
$\nu_e$ (coming from $\nu_\mu\to\nu_e$ oscillation)
interactions, which are classified into
quasi-elastic (qe); $\nu + N \to \ell + N'$,
one pion production (1$\pi$); $\nu + N \to \ell + N' + \pi$,
multi pion production (m$\pi$); $\nu + N \to \ell + N' + n\pi$,
and coherent pion (c$\pi$) production;
$\nu + ^{16}\mbox{O} \to \nu + ^{16}\mbox{O} + \pi^+$,
reactions.
In our computation, we define the signal coming from
the CC interactions in the expected number of events,
${N}_{sig}$,  as follows,
\begin{equation}
{N}_{sig} \equiv  T \sum_{i= qe, 1\pi, m\pi, c\pi}
N_T^i \int dE {\sigma^i}_{CC}(E) \phi^0_{\nu_\mu}(E)
P_{\mu e}(E)  \epsilon(E),
\label{signal}
\end{equation}
where
$N_T^i$ and ${\sigma^i}_{CC}(E)$ are
the the number of target and the CC cross section, respectively,
for $i$-th reaction process,
$T$ is the exposure time,
$\phi^0_{\nu_\mu}(E)$ is $\nu_\mu$ neutrino flux
at the detector site in the absence of oscillation
as a function of neutrino energy,
$P_{\mu e}(E)$ is the
$\nu_\mu \to \nu_e$ oscillation probability,
and $\epsilon(E)$ is the detection efficiency for the
$e$-like events, which was taken from Ref.~\cite{obayashi}. 
See Refs.~\cite{Okumura-Ishihara}, for the neutrino cross 
sections we used in this work. 

We also take into account possible background which
are coming from $\pi^0$ produced in the NC and the CC interactions,
$e/\mu$ misidentification, and
$\nu_e$ contamination in the original $\nu_\mu$ beam,
where the dominant ones come from $\pi^0$ produced in
NC reactions as well as $\nu_e$ contamination.
Following Ref.~\cite{JHF2}, we express background event as
\begin{equation}
N_{BG} = N^{CC}_{BG}  + N^{NC}_{BG}   +  N^{beam}_{BG},
\end{equation}
where $N^{CC}_{BG}$, $N^{NC}_{BG}$
and $N^{beam}_{BG}$ are backgrounds coming from the CC,
the NC interactions and the $\nu_e$ contamination, respectively.
We compute $N^{CC}_{BG}$ in a similar way as
${N}_{sig}$ in Eq. (\ref{signal}) but
$P_{\mu e}(E)$ replaced by $P_{\mu \mu}(E) $
and  $\epsilon(E)$ replaced by a constant
reduction efficiency $\sim 0.017$ \%,
estimated from Table 2 in Ref.~\cite{JHF2},
as an approximation.

Similar to the CC interactions, contributions from
NC ones include the following reactions:
elastic (es); $\nu + N \to \nu + N$,
one pion production; $\nu + N \to \nu + N'+ \pi$,
multi pion production; $\nu + N \to \nu + N'+ n \pi$,
and coherent pion production;
$\nu + ^{16}\mbox{O} \to \nu + ^{16}\mbox{O} + \pi^0$.
We can express $N^{NC}_{BG}$ as
\begin{equation}
N^{NC}_{BG} \equiv \eta T \sum_{i = es, 1\pi, m\pi, c\pi}
N_T^i \int dE {\sigma^i}_{NC}(E) \phi^0_{\nu_\mu}(E),
\label{nc_bg}
\end{equation}
where $\eta$ is the reduction efficiency, approximated as constant,
which is about 0.22 \% of the total NC events, estimated from
Table 2 in Ref.~\cite{JHF2}.

The contribution from the beam $\nu_e$ contamination
can be estimated in a similar way as in
eqs. (\ref{signal}) and  (\ref{nc_bg}) but
$\phi^0_{\nu_\mu}(E)$ replaced by $\phi^0_{\nu_e}(E)$,
the $\nu_e$ flux which exists in the $\nu_\mu$ flux in the absence
of oscillation and $ P_{\mu e}(E)$ by
$ P_{ee}(E)$ in Eq.  (\ref{signal}).

We define the expected number of ``e-like'' events in
$\nu_\mu \to \nu_e$ channel as the sum of signal and
background as,
\begin{equation}
N(e^-) \equiv {N}_{sig} +  N_{BG}.
\end{equation}
For anti-neutrino channel, $\bar{\nu}_\mu \to \bar{\nu}_e$,
we compute the expected number of $e^+$-like events in
the same way properly replacing neutrino flux as well
as cross sections by that of anti-neutrino.

In Table 1 we present expected number of events without
(which means to set $\epsilon = \eta$ =100 \%
in Eqs.(\ref{signal}) and (\ref{nc_bg}))
and with detection efficiency for JHF neutrino OA 2 degree
beam with baseline 295 km and 0.77 MW beam power assuming
Super-Kamiokande detector (22.5 kton),
corresponding to the first phase configuration
of the JHF neutrino project~\cite{JHF2},
with exposure of 5 years.
For comparison we also present the numbers found in new
JHF LOI~\cite{JHF2}.
The numbers we obtained are rather similar to those computed by
the JHF working group, allowing us to have confidence on that
our computations are accurate enough for our purpose.

\vglue 0.5cm
\TABLE{
\caption{\label{tab:number}
Comparison of our expected number of events with that of
the JHF working group for OA 2 degree beam option ~\cite{JHF2}
with baseline 295 km, 0.77 MW beam power and
the exposure of 5 years assuming the SK detector (22.5 kton).
For the calculation of oscillated $\nu_e$ signal,
$\Delta m^2_{13}
 = 3 \times 10^{-3}$ eV$^2$, $\sin^22\theta_{23} = 1.0$,
$\sin^22\theta_{13} = 0.1$, and other mixing parameters are
set to be zero ($\Delta m^2_{12} \to 0$, $\sin^22\theta_{12} \to 0$)
and matter effect was neglected.
For each case, upper and lower number indicate the expected
number of events before reduction (or with 100 \% detection
efficiency) and after reduction (with reduction and/or detection
efficiency).
}
\begin{tabular}{lcccc}
\hline\hline
    & $\nu_\mu$ CC  & $\nu_\mu$ NC & Beam $\nu_e$ & Oscillated $\nu_e$ \\
\hline
This work (before reduction)  & 10707    & 3840     &  270   & 297  \\
This work (after  reduction)  & 1.8      &  8.8     &  10.3  & 142.7  \\
\hline
JHF Phase I (before reduction) & 10714    & 4080      &  292   & 302   \\
JHF Phase I (after  reduction) & 1.8     & 9.3      &  11.1   & 123.2   \\
\hline\hline
\end{tabular}
}


\subsection {Single detector at $L=300$ km}

This is essentially identical with the upgraded JHF neutrino
experiment in its Phase II. While its detailed study is underway
in the working group, we try to make some suggestions here which
might add (we hope) their final proposal some useful ingredients.

In order to have some feelings about the expected number
of events for various JHF neutrino beam options, 
we show in Table 2 the numbers of electron appearance events 
assuming 100 \% $\nu_\mu \to \nu_e$ conversion, for 1 megaton 
SK type detector (with fiducial volume of 0.9 Mton) 
with baseline $L=295$ km, and 
4 MW beam power as planned in the Phase II of the JHF 
neutrino experiment with an exposure of 1 year. 
We also show the background which is essentially
constant and does not depend on neutrino oscillation.
One can estimate the expected number of events
in the presence of neutrino oscillation from the numbers
in the table by simply multiplying oscillation
probability which is properly averaged over the cross section,
neutrino energy spectrum, detection efficiency, etc.

\vglue 0.5cm
\TABLE{
\caption{\label{tab:number}
Expected numbers of $e^-$ and $e^+$-like events in neutrino and
anti-neutrino channels for some possible JHF neutrino beam
options we use in our work, assuming
100 \% conversion of $\nu_\mu \to \nu_e$ and
$\bar{\nu}_\mu \to \bar{\nu}_e$,  respectively. We take a 1 Megaton
SK-type water Cherenkov detector (assuming fiducial volume of 0.9 Mton)
with baseline distance of 295 km, 4 MW beam power
(corresponding to the Phase II of JHF neutrino project),
and exposure of 1 year.
Signal as well as background events are presented separately
in the table.
}
\begin{tabular}{lcccc}
\hline\hline
    &  \multicolumn{2}{c}{Narrow Band Beam}
    & \multicolumn{2}{c}{Off Axis Beam}\\
    & 2 GeV& 3 GeV & 3 deg. & 2 deg. \\
\hline
Peak Energy (GeV)   &   $\sim 1.0$  &  $\sim 1.4$
         &   $\sim 0.55$	 &  $\sim 0.75$ \\
\hline
$\nu$ (signal)    &   $1.8 \times 10^5$ &  $2.7 \times 10^5$
         &   $1.6 \times 10^5$ &  $4.3 \times 10^5$ \\
$\nu$ (background)    &   $1.7 \times 10^2$ &  $2.5 \times 10^2$
         &   $3.5 \times 10^2$ &    $  8.2 \times 10^2$ \\
\hline
$\bar{\nu}$ (signal) & $6.3 \times 10^4$ & $1.2 \times 10^5$
             &  $4.6 \times 10^4$ & $1.4 \times 10^5$ \\
$\bar{\nu}$ (background) & $ 0.6\times 10^2$ & $1.2 \times 10^2$
             &  $ 1.1 \times 10^2$ & $  2.7  \times 10^2$ \\
\hline\hline
\end{tabular}
}

\FIGURE[!ht]{
\centerline{\protect\hbox{
\psfig{file=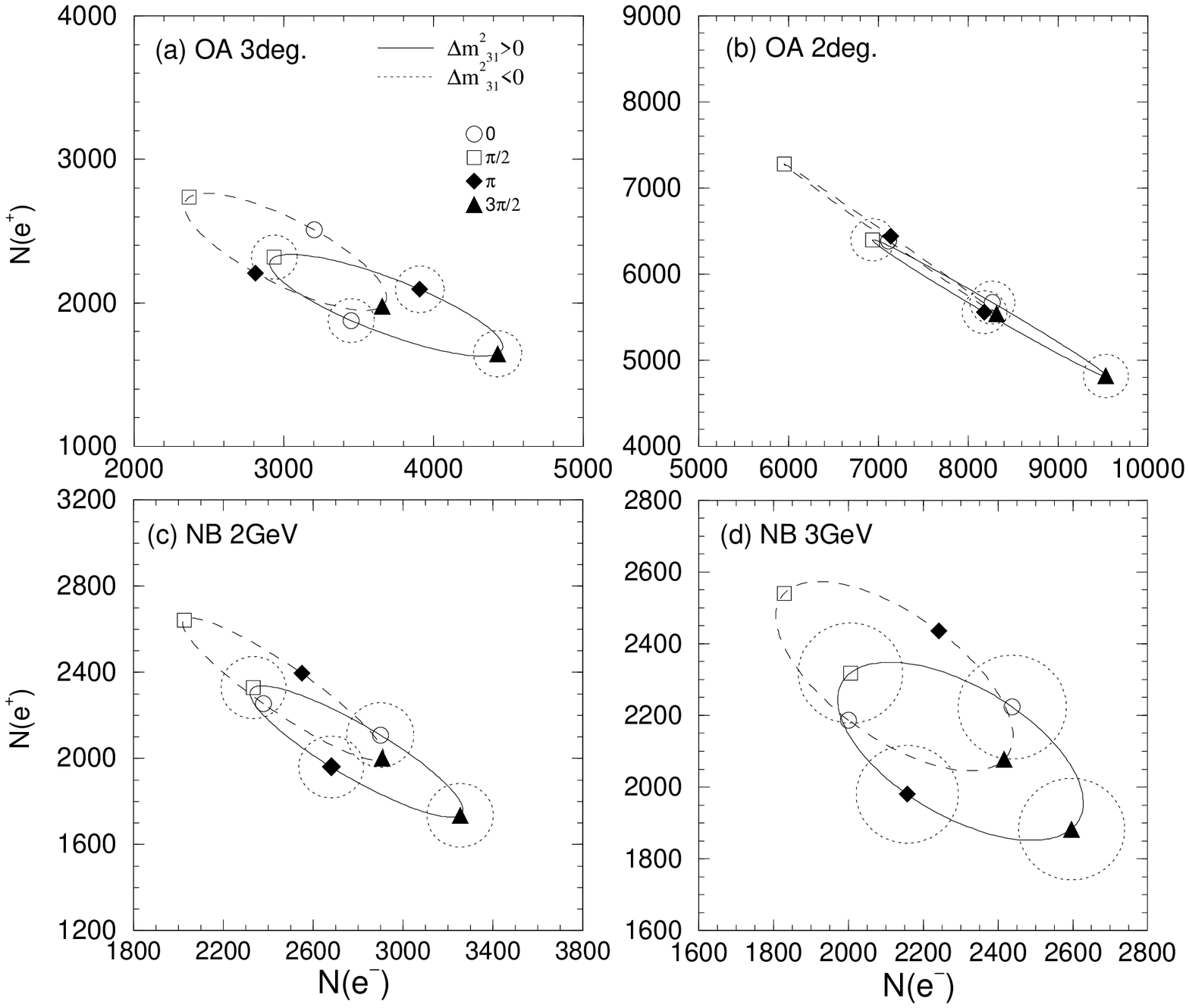,height=15.5cm,width=16.5cm}
}}
\vspace{-1.3cm}
\caption{
CP trajectory in the $N(e^-)-N(e^+)$ plane for
off axis and narrow band beam for $L=295$ km.
Dotted circles indicate 3 $\sigma$ errors
(only statistical ones).
}
\label{Fig6}
}

As we emphasized at the beginning of this section the statistics
heavily depends on the value of $\theta_{13}$, on which we have
no clue at present apart from the CHOOZ constraint \cite {CHOOZ}.
Hence, it is difficult to judge if the experiment that we are
discussing is feasible or not at this moment.
Therefore, we restrict ourselves in this paper to
a particular value of $\theta_{13}$,
$\sin^2{2\theta_{13}} = 0.05$, which is about
a half of the CHOOZ bound.

Now we present in Fig. 6 the CP trajectory contour written
on the plane spanned by numbers of events (not the probabilities),
$N(e^-)$ and $N(e^+)$, including the background assuming 2 and 6
years running for neutrino and anti-neutrino channels, respectively.
The Fig.6a is for OA 3 degree beam which peaks at $E \sim 0.5$ GeV,
while Fig.6b is for OA 2 degree beam which peaks at $E \sim 0.8$ GeV.
Fig.6c and 6d are for NB 2 and 3 GeV beams, whose energy spectra peaks 
at $\sim$ 1 GeV and 1.4 GeV, respectively.
We note that only for NB 3 GeV beam, we assume, as an approximation,
anti-neutrino flux is the same as that of neutrino since
the flux for anti-neutrino was not available for
this particular beam option~\cite{kobayashi}.
Dotted circles in Fig.6 indicate uncertainty which corresponds
to 3 $\sigma$, where only the statistical error is taken into account.
Unfortunately, we have no way of estimating expected systematic errors
in the super-JHF experiments.

As we noticed in Sec. 3 there is an inversion phenomenon in
the diagram. Namely, for a given shape of the contour,
the low probability branch ($\cos{\delta} > 0$ region) in Fig.6a
is mapped into
the high probability branch ($\cos{\delta} < 0$ region) in Fig.6b
in $\Delta m^2_{13} > 0$ case, and
vice versa in $\Delta m^2_{13} < 0$ case.

We point out that the feature can be used as a method of
identifying the value of $\delta$, if otherwise ambiguous in
its measurement. Suppose that they run first at $E=1$ GeV
and obtained the result which tends to prefer the high probability
branch, but they were not confident because the high and the low
probability branches are not so well separated.
While they can just continue to run to increase statistics
with the same energy in such circumstance,
an alternative way (and better way, we believe)
is to run at lower energies, $E \sim 0.5$ GeV.
If the parameter is really in the high (low) probability branch at
$E \sim 1.0$ GeV, then it must jump down (up) to the low (high)
probability branch at $E \sim 0.5$ GeV.

Let us consider, for example, the case of positive $\Delta m^2_{13}$.
If $\delta \simeq \frac{\pi}{4}$ (high probability branch),
the jump is from
$P(\nu) \simeq 2.05$ \% at 1 GeV to $P(\nu) \simeq 1.3$ \% at 0.5 GeV,
as one can see in Fig. 1.
On the other hand, if $\delta \simeq \frac{3\pi}{4}$ (low probability
branch) the jump is from
$P(\nu) \simeq $ 1.85\% at 1 GeV to $P(\nu) \simeq 1.65$ \% at 0.5 GeV.
Thus, a greater downward jump of about 37 \% is predicted in the case
of $\delta \simeq \frac{\pi}{4}$,
as compared to a modest $\sim$ 10 \% decrease in the
$\delta \simeq \frac{3\pi}{4}$ case.
Thus, running at high and low energies would greatly help to distinguish
the high and the low probability branches,
or $\cos{\delta} > 0$ and $\cos{\delta} < 0$ cases,
if the energies are tuned to have a large radial thickness.
Unfortunately, we cannot resolve the two-fold ambiguity by using
this technique, because the high probability branch jumps to
low probability branch, or {\it vise veasa}, both in
$\Delta m^2_{13} > 0$ and
$\Delta m^2_{13} < 0$ cases simultaneously.

It should be emphasized that, if we are lucky, we will be able
to determine the CP violating angle $\delta$ and the sign of
$\Delta m^2_{13}$ simultaneously. It is the case with highest
statistics among the cases considered in this paper, and hence
it is likely to have resolving power of the both quantities.
By luckiness, we mean very roughly that if
the angle $\delta$ is in the third or the fourth quadrants
in the case of normal mass hierarchy ($\Delta m^2_{13} > 0$),
and is the first or the second quadrants
in the case of inverted mass hierarchy ($\Delta m^2_{13} < 0$),
respectively.
In this case, the experiment will reveal by high statistics data
taking the CP violating angle and the sign of $\Delta m^2_{13}$
simultaneously. The accuracy of the measurement of $\delta$ would
be somewhere in 20-30 degrees, if the systematic errors are not
too sizable.

\subsection {Single detector at $L=700/1000$ km}

\FIGURE[!ht]{
\centerline{\protect\hbox{
\psfig{file=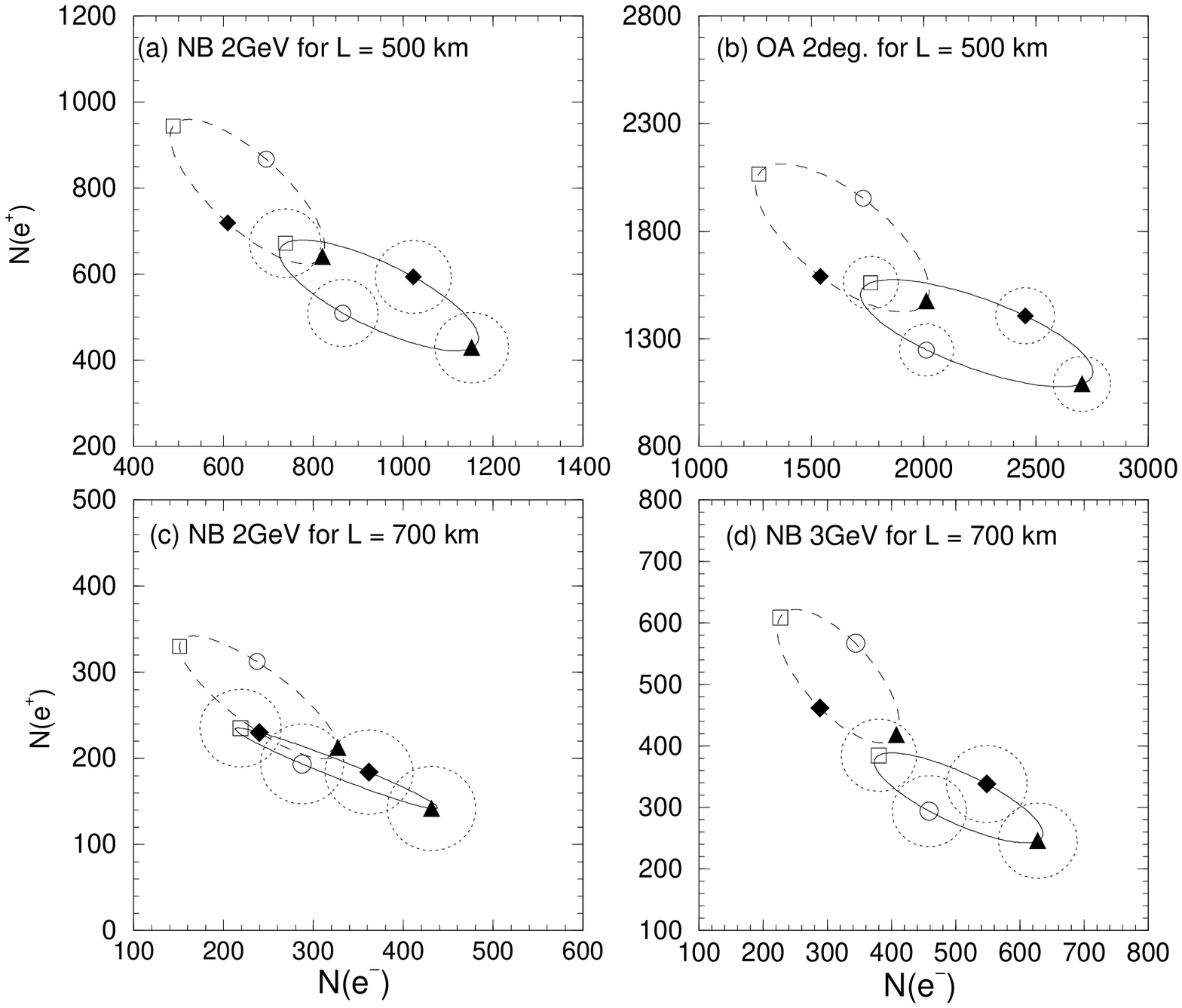,height=15.5cm,width=16.5cm}
}}
\vspace{-1.3cm}
\caption{
CP trajectory in the $N(e^-)-N(e^+)$ plane for
off axis and narrow band beam for $L=500$ km or
700 km.
Dotted circles indicate 3 $\sigma$ errors
(only statistical ones).
Mixing parameters are fixed to be the same as in Fig. 1.
}
\label{Fig7}
}
We next consider the option of single detector at $L=700/1000$ km.
As we saw in Sec. 3 the detector at $L=700/1000$ km shows a
qualitatively different characteristics;
the neutrino and the antineutrino trajectory becomes nonoverlapping
for neutrino energies of $1-2$ GeV, allowing simultaneous
determination of $\delta$ and the sign of $\Delta m^2_{13}$
for any values of $\delta$, if the statistics is high enough.

In Fig. 7 we present the CP trajectory contours written on the
$N(e^-) - N(e^+)$ plane with path length of (a, b) 500 km and
(c, d) 700 km. The former is for comparison, in particular to
represent the correlation between the matter effect and the number
of events as a function of baseline length. Since there
is no version of off-axis beam with peak energies of $\gsim 1$ GeV
available to us, we use NB 3 GeV beam which peaks at around $E=1.4$ GeV.
It appears that off-axis beams are more intense than narrow-band beams
with similar peak energies by a factor of 2-3.
If we are able to design OA beam with peak energy of
$\gsim 1.4$ GeV, we would gain a factor of 3 in number of events.
If it is the case, we may scale the abscissa and the ordinate
by factor of 3 and the radii of the error circle are reduced by
factor of $\sqrt{3}$.
Of course, to have off-axis beam with peak energies of $\gsim 1$ GeV
one should probably start with a completely different proton beam
design with higher energies, whose discussion is far beyond the
scope of this paper.

We note that for NB 3 GeV the number of events is $\sim$ 500 for
2 (6) years of running in neutrino (antineutrino) channel at
$L = 700$ km, and $\sim$ 2000 for $L = 500$ km. They are certainly
not enormous but are not too small either. If the beam normalization
is properly known it would be possible to discover the existence of
leptonic CP violation and at the same time to determine the sign of
$\Delta m^2_{13}$.

Clearly, the conclusion depends on many unknown factors, the values
of mixing parameters, $\theta_{13}$, $\Delta m^2_{12}$, and $\delta$
and also one experimental conditions, uncertainties in absolute flux
normalization, cross sections, and efficiency background rejection.

\subsection {Two detectors at $L=300$ and $700/1000$ km}

Suppose that we have a bad luck in the experiment with $L=300$ km
by having $\delta$ in the alternative regions from that we have
mentioned in Subsec. 6.2. Namely, if
$\sin{\delta} > 0$ in the case of normal mass hierarchy and
$\sin{\delta} < 0$ in the case of inverted mass hierarchy,
we will be able to determine (assuming enough statistics)
$\delta$ but only with modulo two-fold ambiguity.
While it is already a great achievement, it may be better if we
have a way to resolve the ambiguity.
Motivated by the consideration in the preceding section, it is
natural to consider the two-detector option, one at $L=300$ km and the
other at $L=700/1000$ km. While it is indeed an ``expensive option''
which uses the two megaton detectors
(or possibly one 1 Mton and one $\sim$ 100 kton iron calorimeter)
it might not be an unrealistic one in view of the proposal
of detectors either in Korea \cite{korea}, or in Beijin \cite {beijin}.

In this two-detector option, the requirement of statistics can
be relaxed because the second detector is effectively only for
the determination of the sign of $\Delta m^2_{13}$.
In view of Fig. 7 a narrow band beam with energy a bit higher than
NB 3 GeV beam may be able to do the job, without multiplying 3 in
flux normalization

\section {CP trajectory diagram for neutrino factories}

Finally, we briefly address in this section the features of CP
trajectory diagram for physical parameters which are appropriate
for neutrino factories.
It is to illustrate enormous difference in the features of the
CP phase-matter interplay between the situations of
neutrino factory and the low energy superbeam.
We neither intend to make full comparison of these two different
strategies nor try to argue which is the better way to measure
CP violation.

\FIGURE[!ht]{
\centerline{\protect\hbox{
\psfig{file=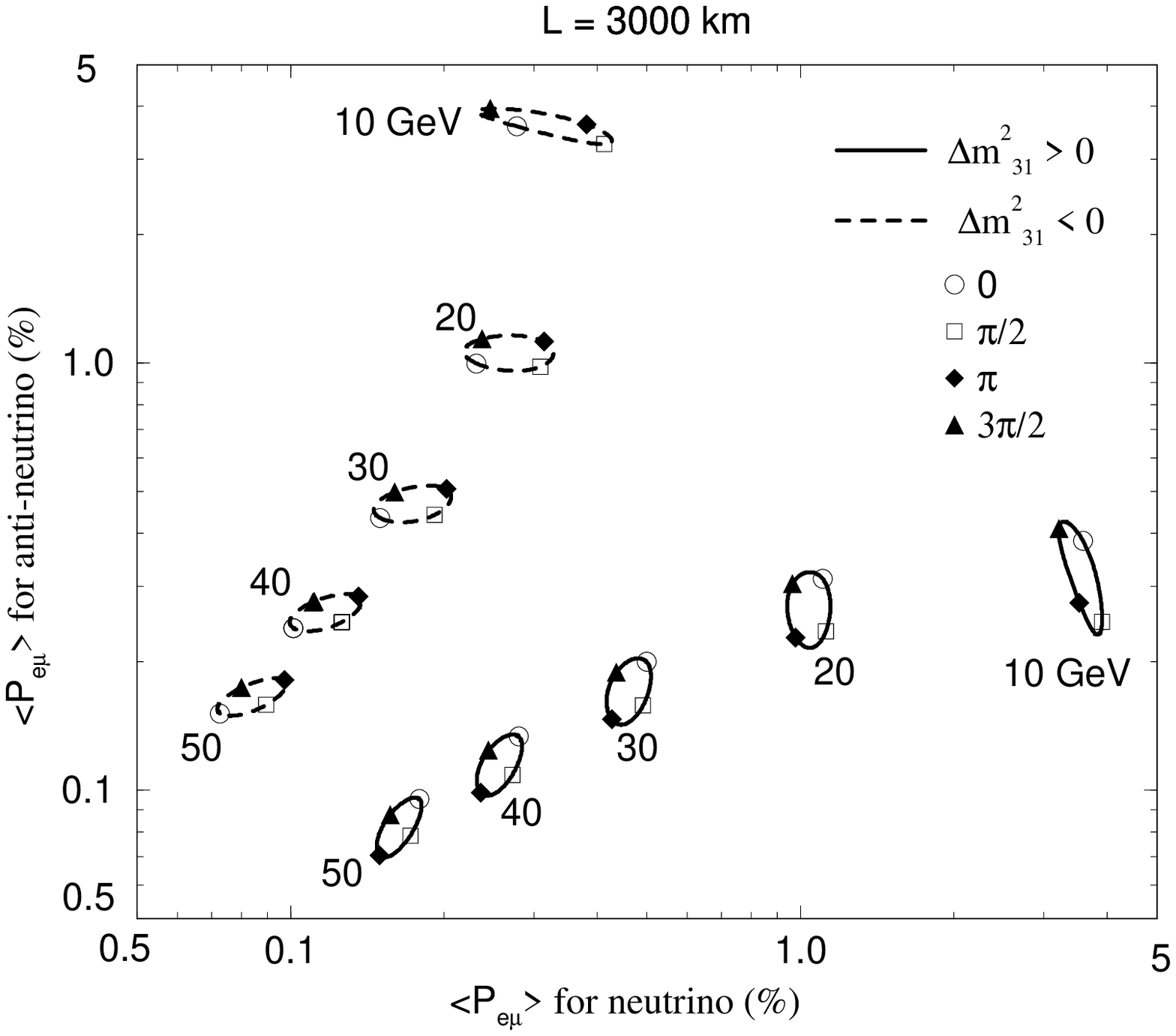,height=15.cm,width=16.cm}
}}
\vspace{-1.6cm}
\caption{
The CP trajectory diagram in bi-probability plane for $L=3000$ km
and much higher neutrino energies $E = 10-50$ GeV which correspond
to so called ``Neutrino Factory'' situation.
The mixing parameters are fixed to be the same
as in Fig. 1 except that we take $\rho Y_e = 2.0$ g/cm$^3$.
}
\label{Fig8}
}

In Fig. 8, presented is the CP trajectory diagram for $L = 3000$ km
for varying neutrino energies from 10 GeV to 50 GeV.
Only the trajectories with matter effect are plotted for both
sign of $\Delta m^2_{13}$. The matter density is taken as
$\rho Y_e = 2$ g/cm$^3$ corresponding to upper mantle region.
Unlike diagrams for low energy beams presented in the preceding
sections the two trajectories corresponding to positive and
negative $\Delta m^2_{13}$ are far apart with each other.
It leads to the well known fact that the matter effect dominates
over the CP violating effect due to the Kobayashi-Maskawa phase.

The good news is that in the neutrino factory situation there is
of course no such ambiguity as the one we discussed in this paper.
(However, there arises an another ambiguity in this case, if
the value of $\theta_{13}$ is not known, as pointed out by
Burguet-Castell et al. \cite {nufact.vs.sb}, as mentioned in
\cite {2detector}.)
On the other hand, the problem of matter effect contamination
is severer.
It appears to us that we need further studies to clearly
separate genuine CP violating effect from the matter effect in
neutrino factories.

\newpage
\section {Conclusion}

We have discussed the features of the neutrino oscillations
which are relevant for the experiments with conventional
low energy (typically $\sim $ 1 GeV) superbeams with mediumly-long
baseline ($\lsim 1000$ km) which aim at measuring leptonic CP violation.
We have assumed, for the purpose of
estimating number of events and statistical uncertainties,
a supermassive 1 megaton water Cherenkov detector of
Super-Kamiokande type, and 4 MW beam power as planned in the
Phase II of the JHF neutrino experiment.

We started with focusing on the problem of interplay between genuine
CP violation and the matter effect. While it is a widely
discussed topics in the literature, we have uncovered an
interesting new feature; the matter effect helps.
Namely, the matter effect in such mediumly-long baseline experiments
is comfortably large so that it helps to resolve the (approximate)
two-fold ambiguity which would exist, in the absense of the effect,
in the vacuum oscillation probability.

To elucidate the features of the CP phase-matter interplay,
we have introduced a new powerful tool called the ``CP trajectory
diagram in bi-probability space''.
The diagram allows us to represent pictorially the following three
effects separately in a single diagram in a compact form:
1) the genuine CP violating effect due to the leptonic
Kobayashi-Maskawa phase $\delta$, or $\sin\delta$, the CP odd term,
2) the effect due to the CP even term, $\cos\delta$,
and
3) the fake CP violating effect due to the
earth matter.
As discussed in the text,
the effect of the CP odd term is characterized by
the ``polar thickness'' whereas the effect of the CP even term is
characterized by the ``radial thickness''
of the trajectory contour
(see the beginning of Sec. 4 for the definitions
of ``polar'' and ``radial'' thinkness).
On the other hand, the earth matter effect
is characterized by the distance of separation between
the two trajectory contours with different
sign of $\Delta m^2_{13}$.

By using such diagram, we have observed that there is
a two-fold ambiguity in the determination of the CP violating
angle $\delta$, if we do not know {\it a priori} the sign of
$\Delta m^2_{13}$.
The ambiguity is shown to be a remnant of the approximate
symmetry possessed by the vacuum oscillation
probability (as mentioned above) under simultaneous transformation
of $\delta \rightarrow \pi - \delta$ and
$\Delta m^2_{13} \rightarrow - \Delta m^2_{13}$.
We have discussed the principle of "maximal fatness"
of the CP trajectory diagram to find optimal experimental
parameters in order to have a large CP violation and at
the same time to resolve such two-fold ambiguity.

By these considerations we are naturally invited to consider
the possibility of measuring simultaneously the CP phase $\delta$
as well as the sign of $\Delta m^2_{13}$.
We have discovered the enlighting possibility that such simultaneous
measurement can be done with relatively short baseline,
$L \simeq 300$ km assuming 1 megaton water Cherenkov detector,
as in the phase II of JHF-Kamioka neutrino experiment.
It works, however, only under the condition of nature's kind setting
of the parameter into the region $\sin{\delta} \cdot \Delta m^2_{13} < 0$.

We also described a way of resolving the degeneracy in the case of
unresolved high and the low probability branches, or
$\cos{\delta} > 0$ and $\cos{\delta} < 0$ region
for $\Delta m^2_{13} > 0$ case, and
vice versa in $\Delta m^2_{13} < 0$ case.
By running at high ($\sim 1$ GeV) and low ($\sim 0.5$ GeV)
beam energies the high and low probability branches are interchanged
with each other, and this effect may help in identifying $\delta$
while it cannot resolve the two-fold ambiguity.

Furthermore, we have uncovered an interesting possibility that
simultaneous measurement of $\delta$ and the sign of $\Delta m^2_{13}$
can be done {\it in situ} for the whole region of $\delta$
($0 < \delta < 2 \pi$) in the experiments with longer baseline
distance, $L \sim 700$ km (corresponding to the CERN to Gran Sasso
and/or Fermilab to Soudan-2 distances), again with a
megaton water Cherenkov detector.

We also briefly discussed how the CP trajectory diagram look
like for the neutrino factory situation and observed that
the two CP trajectory with different signs of $\Delta m^2_{13}$
are well separated owing to the fact that the matter effect is
dominant. It is in sharp contrast with the case of low energy
superbeam, the main subject of this paper.
A clear way of separating the CP phase effect from the matter effect
is naturally called for.

We hope that our discussions in this paper would help in designing
the future neutrino oscillation experiments.

\acknowledgments

We have been greatly benefited by massive communications
with members of the JHF-SK Neutrino Working Group, in particular,
Akira Konaka, Takashi Kobayashi,
and most intensively, Yoshihisa Obayashi
for numerous valuable informative correspondences on low energy
neutrino beams and detector backgrounds, and sharing with us
many preliminary informations of the JHF experiments
which are prepared for the new version of the Letter of Intent
that appeared quite recently.

One of us (HM) thanks Marcelo M. Guzzo and the Department of
Cosmic Ray and Chronology at Gleb Wataghin Physics Institute
in UNICAMP for their hospitality during his visit where part
of this work was done.
This work was supported by the Brazilian funding agency
Funda\c{c}\~ao de Amparo \`a Pesquisa do Estado de S\~ao Paulo (FAPESP),
and by the Grant-in-Aid for Scientific Research in Priority Areas
No. 11127213, Japan Ministry of Education, Culture, Sports, Science
and Technology.

\vskip 1.0cm

\begin{center}
{\large APPENDIX}
\end{center}

We give a proof that the CP trajectory is elliptic in this Appendix.
It can be shown under the assumption of mass ($\Delta m^2$) hierarchy
and the adiabatic approximation that the neutrino
and the antineutrino oscillation probabilities can be written in
the following forms:
\begin{eqnarray}
P(\nu) &=& A \cos{\delta} + B \sin{\delta} + C \\
P(\bar{\nu}) &=& \bar{A} \cos{\delta} - \bar{B} \sin{\delta} + \bar{C}
\label{general}
\end{eqnarray}
It is true in all the perturbative formula so far derived in
Refs. \cite {AKS97,MN97,MN98,recent}.
However, we should note that it is not true in the exact formula
derived in \cite {ZS88}. Therefore, it is not an exact formula
apart from the case in vacuum, but is a very good one in the case
of hierarchical mass difference we are interested in.

Once the above general form holds for a fixed arbitrary energy,
it hold even if one take average over an arbitrary neutrino
energy spectrum. Therefore, we implicitly imply
$P$'s and the coefficients in (\ref{general}) as averaged over
an energy spectrum, though it is not indicated explicitly.
Hence, the general form applies to the CP trajectory diagram on
bi-number-of-event plane, with which we have dealt in Sec. 6.
(In fact, this general form applies even if we take average over
the neutrino path length.)

By eliminating $\delta$ we obtain the equation obeyed by
$P(\nu)$ and $P(\bar{\nu})$ as

\begin{equation}
\frac{1}{\left(\frac{A}{B} + \frac{\bar{A}}{\bar{B}}\right)^2}
\left(\frac{P(\nu) - C}{B} +
\frac{P(\bar{\nu}) - \bar{C}}{\bar{B}}\right)^2 +
\frac{1}{\left(\frac{B}{A} + \frac{\bar{B}}{\bar{A}}\right)^2}
\left(\frac{P(\nu) - C}{A} -
\frac{P(\bar{\nu}) - \bar{C}}{\bar{A}}\right)^2
= 1.
\label{ellip.mat}
\end{equation}
In vacuum, the expression simplifies because $A=\bar{A}$ etc.:
\begin{equation}
\left(\frac{P(\nu) + P(\bar{\nu}) - 2C}{2 A}\right)^2 +
\left(\frac{P(\nu) - P(\bar{\nu})}{2 B}\right)^2 = 1,
\label{ellip.vac}
\end{equation}
which implies that the minor (if $A > B$) or major (if $A < B$) axes are
always at 45 degree.

This completes the proof that the CP trajectory is elliptic.

%
%



\begin{thebibliography}{99}

\bibitem {SKatm}
Kamiokande Collaboration, Y.~Fukuda {\it et al.},
{\it Atmospheric muon-neutrino/electron-neutrino
ratio in the multi GeV energy range,
Phys.\ Lett.\ }B {\bf 335} (1994) 237;\\
%
Super-Kamiokande Collaboration, Y.~Fukuda  {\it et al.},
{\it Evidence for oscillation of atmospheric neutrinos,
Phys. Rev. Lett. }{\bf 81} (1998) 1562;\\
%
S.~Fukuda  {\it et al.},
{\it Tau neutrinos favored over sterile neutrinos in atmospheric
muon  neutrino oscillations, ibid. } {\bf 85} (2000) 3999;\\
%
T. Kajita for Super-Kamiokande and Kamiokande Collaboration,
{\it
Atmospheric neutrino results from Super-Kamiokande and Kamiokande:
Evidence for nu/mu oscillations},
in {\it Neutrino Physics and Astrophysics},
{\it Proceedings of the XVIIIth International Conference on Neutrino
Physics and Astrophysics (Neutrino '98)}, June 4-9, 1998, Takayama,
Japan, edited by Y. Suzuki and Y. Totsuka,
(Elsevier Science B.V., Amsterdam, 1999) page 123
[{\it Nucl.\ Phys.\ Proc.\ Suppl.}\  {\bf 77} (1999) 123].

\bibitem {solar}
Homestake Collaboration, K.~Lande {\it et al.},
{\it Measurement of the solar electron neutrino flux with
the Homestake  chlorine detector,
Astrophys.\ J.\ } {\bf 496} (1998) 505;\\
%
SAGE Collaboration, J.~N.~Abdurashitov {\it et al.},
{\it Measurement of the solar neutrino capture rate with gallium metal,
Phys.\ Rev.\ }C {\bf 60} (1999) 055801;\\
%
GALLEX Collaboration, W.~Hampel {\it et al.},
{\it GALLEX solar neutrino observations: Results for GALLEX IV,
Phys.\ Lett.\ } B {\bf 447} (1999) 127;\\
%
Kamiokande Collaboration, Y.~Fukuda {\it et al.},
{\it Solar neutrino data covering solar cycle 22,
Phys.\ Rev.\ Lett.\  }{\bf 77} (1996) 1683;\\
%
Super-Kamiokande Collaboration,
Y.~Fukuda {\it et al.},
{\it Measurements of the solar neutrino flux from
Super-Kamiokande's first  300 days,
Phys.\ Rev.\ Lett.\  }{\bf 81} (1998) 1158
[{\it Erratum-ibid.}\  {\bf 81} (1998) 4279];\\
%
{\it Measurement of the solar neutrino energy spectrum
using neutrino  electron scattering,
ibid.} {\bf 82} (1999) 2430;\\
%
{\it Constraints on neutrino oscillation parameters
from the measurement of  day-night solar neutrino fluxes
at Super-Kamiokande,
ibid.}\  {\bf 82} (1999) 1810;\\
%
S.~Fukuda {\it et al.},
{\it Solar $^8$B and hep neutrino measurements from 1258
days of  Super-Kamiokande data,
Phys.\ Rev.\ Lett.}\  {\bf 86} (2001) 5651;\\
%
{\it Constraints on neutrino oscillations using
1258 days of  Super-Kamiokande solar neutrino data,
ibid.}\  {\bf 86} (2001) 5656.

\bibitem {SNO}
SNO Collaboration, Q.~R.~Ahmad {\it et al.},
{\it Measurement of the rate of $\nu_e+d\rightarrow p+p+e^-$
interactions produced by $^8$B solar neutrinos at the Sudbury
Neutrino Observatory}
nucl-ex/0106015,
see also http://www.sno.phy.queensu.ca/sno/first\_results/.

\bibitem {SKsolar}
Super-Kamiokande Collaboration,  Y.\ Fukuda {\it et al.},
in Ref.~\cite{solar}.

\bibitem {K2K}
K2K Collaboration, S.~H.~Ahn {\it et al.},
{\it Detection of accelerator produced neutrinos at a distance of 250-km,
Phys.\ Lett.}\ B {\bf 511} (2001) 178;\\
See also http://neutrino.kek.jp/news/2001.07.10.News/index-e.html.


\bibitem {MNS}
Z.~Maki, M.~Nakagawa and S.~Sakata,
{\it Remarks on the unified model of elementary particles,
Prog.\ Theor.\ Phys.}\  {\bf 28} (1962) 870.

\bibitem {KM}
M.~Kobayashi and T.~Maskawa,
{\it CP violation in the renormalizable theory of weak interaction,
Prog.\ Theor.\ Phys.}\  {\bf 49} (1973) 652.

\bibitem {early}
The early references on leptonic CP violation may be found in:
N.~Cabibbo,
{\it Time reversal violation in neutrino oscillation,
Phys.\ Lett.}\ B {\bf 72} (1978) 333;\\
%
V.~Barger, K.~Whisnant and R.~J.~Phillips,
{\it CP violation in three neutrino oscillations,
Phys.\ Rev.\ Lett.}\  {\bf 45} (1980) 2084;\\
%
S. Pakvasa, in {\it Proceedings of the XXth International Conference
on High Energy Physics}, edited by L. Durand and L. G. Pondrom,
AIP Conf. Proc. No. 68 (AIP, New York, 1981), Vol. 2, pp. 1164.

\bibitem {NOW2000mina}
H.~Minakata,
{\it The three neutrino scenario,
Nucl.\ Phys.\ Proc.\ Suppl.}\  {\bf 100} (2001) 237
[hep-ph/0101231].

\bibitem {CHOOZ}
CHOOZ Collaboration, M.~Apollonio {\it et al.},
{\it Initial results from the CHOOZ long baseline
reactor neutrino  oscillation experiment,
Phys.\ Lett.}\ B {\bf 420} (1998) 397;\\
%
{\it Limits on neutrino oscillations from the CHOOZ experiment,
ibid.} B {\bf 466} (1999) 415;\\
See also, The Palo Verde Collaboration,
F.~Boehm {\it et al.},
{\it Results from the Palo Verde neutrino oscillation experiment,
Phys.\ Rev.}\ D {\bf 62} (2000) 072002.


\bibitem {JHF}
JHF Neutrino Working Group, Y. Itow {\it et al.},
{\it Letter of Intent:
A Long Baseline Neutrino Oscillation Experiment
Using the JHF 50 GeV Proton-Synchrotron
and the Super-Kamiokande Detector}, February 3, 2000.

\bibitem {MINOS}
The MINOS Collaboration, P. Adamson  {\it et al.},
{\it MINOS Detectors Technical Design Report, Version 1.0},
NuMI-L-337, October 1998.

\bibitem {OPERA}
OPERA Collaboration, M. Guler {\it et al.},
{\it OPERA: An Appearance Experiment to Search for Nu/Mu
$\leftarrow$$\rightarrow$ Nu/Tau
Oscillations in the CNGS Beam. Experimental Proposal},
CERN-SPSC-2000-028, CERN-SPSC-P-318, LNGS-P25-00, Jul 2000.

\bibitem {MN01}
H.~Minakata and H.~Nunokawa,
{\it Inverted hierarchy of neutrino masses disfavored
by supernova 1987A,
Phys.\ Lett.}\ B {\bf 504} (2001) 301
[hep-ph/0010240].

\bibitem{MSW}
S.~P.~Mikheev and A.~Y.~Smirnov,
{\it Resonant amplification of neutrino oscillations
in matter and solar neutrino spectroscopy,
Nuovo Cim.}\ C {\bf 9} (1986) 17;\\
%
L.~Wolfenstein,
{\it Neutrino oscillations in matter,
Phys.\ Rev.}\ D {\bf 17} (1978) 2369.


\bibitem {MN00}
H.~Minakata and H.~Nunokawa,
{\it Measuring leptonic CP violation by low energy
neutrino oscillation  experiments,
Phys.\ Lett.}\ B {\bf 495} (2000) 369.


\bibitem {Nufact00nuno}
H.~Minakata and H.~Nunokawa,
{\it Measuring CP violation by low-energy medium-baseline
neutrino  oscillation experiments},
Talk given at International Workshop on Neutrino Factories
based on Muon Storage Rings (NuFACT00),
Monterey, California, 22-26, May 2000, hep-ph/0009091.

\bibitem {richter}
B.~Richter,
{\it Conventional beams or neutrino factories:
The next generation of  accelerator-based neutrino experiments},
hep-ph/0008222.

\bibitem {mimicking}
Some recent discussions of the vacuum mimicking mechanism may be
found in \cite {NOW2000mina} and in
%
S.~J.~Parke and T.~J.~Weiler,
{\it Optimizing T-violating effects for neutrino oscillations in matter,
Phys.\ Lett.}\ B {\bf 501} (2001) 106;\\
%
P.~Lipari,
{\it CP violation effects and high energy neutrinos,
Phys.\ Rev.}\ D {\bf 64} (2001) 033002;\\
%
O.~Yasuda,
{\it Vacuum mimicking phenomena in neutrino oscillations},
hep-ph/0106232.

\bibitem {AKS97}
J.~Arafune and J.~Sato,
{\it CP and T violation test in neutrino oscillation,
Phys.\ Rev.}\ D {\bf 55} (1997) 1653;\\
%
J.~Arafune, M.~Koike and J.~Sato,
{\it CP violation and matter effect in long baseline
neutrino oscillation  experiments,
Phys.\ Rev.}\ D {\bf 56} (1997) 3093
[{\it Erratum-ibid.}\ D {\bf 60} (1997) 119905].

\bibitem {MN97}
H.~Minakata and H.~Nunokawa,
{\it How to measure CP violation in neutrino oscillation experiments?,
Phys.\ Lett.}\ B {\bf 413} (1997) 369.

\bibitem {MN98}
H.~Minakata and H.~Nunokawa,
{\it CP violation vs. matter effect in long-baseline
neutrino oscillation  experiments,
Phys.\ Rev.}\ D {\bf 57} (1998) 4403.

\bibitem {recent}
M.~Tanimoto,
{\it Is CP violation observable in long baseline
neutrino oscillation  experiments?,
Phys.\ Rev.}\ D {\bf 55} (1997) 322;\\
{\it Prediction on CP violation in long baseline neutrino
oscillation  experiments,
Prog.\ Theor.\ Phys.}\  {\bf 97} (1997) 901;\\
%
S.~M.~Bilenkii, C.~Giunti and W.~Grimus,
{\it Long-baseline neutrino oscillation experiments
and CP violation in the  lepton sector,
Phys.\ Rev.}\ D {\bf 58} (1998) 033001;\\
%
O.~Yasuda,
{\it Three flavor neutrino oscillations and
application to long baseline  experiments,
Acta Phys.\ Polon.}\ B {\bf 30} (1999) 3089;\\
%
M.~Koike and J.~Sato,
{\it CP and T violation in long baseline experiments
with low energy  neutrino from muon storage ring,
Phys.\ Rev.}\ D {\bf 61} (2000) 073012
[{\it Erratum-ibid.}\ D {\bf 62} (2000) 079903];\\
%
K.~Dick, M.~Freund, M.~Lindner and A.~Romanino,
{\it CP-violation in neutrino oscillations,
Nucl.\ Phys.}\ B {\bf 562} (1999) 29;\\
%
T.~Ota and J.~Sato,
{\it Matter profile effect in neutrino factory,
Phys.\ Rev.}\ D {\bf 63} (2001) 093004;\\
%
T.~Miura, E.~Takasugi, Y.~Kuno and M.~Yoshimura,
{\it The matter effect to T-violation at a neutrino factory,
Phys.\ Rev.}\ D {\bf 64} (2001) 013002.
%
%

\bibitem {cadenas}
J.~J.~Gomez-Cadenas {\it et al.}  [CERN working group on Super Beams
                  Collaboration],
{\it Physics potential of very intense conventional neutrino beams},
Talk at 9th International Workshop on
Neutrino Telescopes, Venice, March 6-9, 2001;
hep-ph/0105297.

\bibitem {nufact1}
A.~De Rujula, M.~B.~Gavela and P.~Hernandez,
{\it Neutrino oscillation physics with a neutrino factory,
Nucl.\ Phys.}\ B {\bf 547} (1999) 21;\\
%
A.~Blondel {\it et al.},
{\it The neutrino factory: Beam and experiments,
Nucl.\ Instrum.\ Meth.}\ A {\bf 451} (2000) 102;\\
%
A.~Cervera, A.~Donini, M.~B.~Gavela, J.~J.~Gomez Cadenas,
P.~Hernandez, O.~Mena and S.~Rigolin,
{\it Golden measurements at a neutrino factory,
Nucl.\ Phys.}\ B {\bf 579} (2000) 17
[{\it Erratum-ibid.}\ B {\bf 593} (2000) 731];\\
%
A.~Donini, M.~B.~Gavela, P.~Hernandez and S.~Rigolin,
{\it Four species neutrino oscillations at nu-factory:
Sensitivity and  CP-violation,
Nucl.\ Instrum.\ Meth.}\ A {\bf 451} (2000) 58.

\bibitem {nufact2}
V.~Barger, S.~Geer and K.~Whisnant,
{\it Long baseline neutrino physics with a muon
storage ring neutrino  source,
Phys.\ Rev.}\ D {\bf 61} (2000) 053004;\\
%
V.~Barger, S.~Geer, R.~Raja and K.~Whisnant,
{\it Long-baseline study of the leading neutrino
oscillation at a neutrino  factory,
Phys.\ Rev.}\ D {\bf 62} (2000) 013004;\\
%
{\it Neutrino oscillations at an entry-level neutrino factory and beyond,
Phys.\ Rev.}\ D {\bf 62} (2000) 073002;\\
%
{\it Short-baseline neutrino oscillations at a neutrino factory,
Phys.\ Rev.}\ D {\bf 63} (2001) 033002;\\
%
C.~Albright {\it et al.},
{\it Physics at a neutrino factory},
hep-ex/0008064.
%

\bibitem {KS00}
A neutrino factory version of low energy experiment is discussed by
M.~Koike and J.~Sato in Ref.~\cite{recent}.

\bibitem {nufact.vs.sb}
V.~Barger, S.~Geer, R.~Raja and K.~Whisnant,
{\it Exploring neutrino oscillations with superbeams,
Phys.\ Rev.}\ D {\bf 63} (2001) 113011;\\
%
V.~Barger {\it et al.},
{\it Oscillation measurements with upgraded conventional neutrino beams},
hep-ph/0103052;\\
%
K.~Dick, M.~Freund, P.~Huber and M.~Lindner,
{\it A comparison of the physics potential of future
long baseline neutrino  oscillation experiments,
Nucl.\ Phys.}\ B {\bf 598} (2001) 543;\\
%
M.~Koike, T.~Ota and J.~Sato,
{\it Ambiguities of theoretical parameters and
CP/T violation in neutrino  factories},
hep-ph/0011387;\\
%
J.~Burguet-Castell, M.~B.~Gavela, J.~J.~Gomez-Cadenas,
P.~Hernandez and O.~Mena,
{\it On the measurement of leptonic CP violation},
hep-ph/0103258;\\
%
J.~Pinney and O.~Yasuda,
{\it Correlations of errors in measurements of
CP violation at neutrino  factories},
hep-ph/0105087.

\bibitem {kobayashi}
T. Kobayashi, Talk at 5th TOKUTEI-RCCN Workshop, Institute for
Cosmic Ray Research, Chiba, Japan, February 23-24, 2001,
and private communications.


\bibitem {obayashi}
Y. Obayashi,  Talk at Workshop on JHF-SK Neutrino Oscillation Experiment,
April 3-4, 2001, KEK, Tsukuba, Japan and private communications.


\bibitem {JHF2}
Y. Itow {\it et al.},
{\it The JHF-Kamioka neutrino project}, hep-ex/0106019.

\bibitem {mina.nufact01}
Part of the results in this and the following sections has been
announced; H. Minakata, Talk at The 3rd International Workshop on
Neutrino Factories based on Muon Strage Rings (NuFACT'01),
May 24-30, 2001, Tsukuba, Japan. See,
http://www-prism.kek.jp/nufact01/May29/WG1/29wg1-(underline)minakata.pdf.

\bibitem {FLMP}
G.~L.~Fogli, E.~Lisi, D.~Montanino and A.~Palazzo,
{\it Three-flavor MSW solutions of the solar neutrino problem,
Phys.\ Rev.}\ D {\bf 62} (2000) 013002.

\bibitem {2detector}
In the Brookhaven proposal
of long baseline neutrino oscillation experiment the two-detector
method is advocated to have increased sensitivity
in measurement of the disappearance effect. See
%
D. Beavis  {\it et al.} (E889 Collaboration),
Long Baseline Neutrino Oscillation Experiment at the AGS Approved
by the HENPAC as AGS Experiment 889,
Physics Design Report, BNL No. 52459, April 1995.
%
We have discussed the two- and three-detector methods
to extract the CP violating piece of the neutrino oscillation
probabilities without doing antineutrino (or neutrino) experiment
\cite {MN97}.
Very recently, Burguet-Castell et al. in \cite {nufact.vs.sb}
discussed two-detector method for resolving a similar two-fold
ambiguity in the simultaneous determination of $\delta$ and
$\theta_{13}$ in neutrino factory.


\bibitem {konaka.strategy}
It would be very interesting to re-examine the problem of how to
optimize the experimental parameters with use of informaton of
neutrino energy obtained by reconstrucion of events in an
event-by-event basis, as pointed out by A. Konaka through
private communications.


\bibitem {shiozawa}
M. Shiozawa, Talk at Workshop on JHF-SK Neutrino Oscillation Experiment,
April 3-4, 2001, KEK, Tsukuba, Japan.

\bibitem {Okumura-Ishihara}
K. Okumra, Doctor thesis, 
{\it Observation of atmospheric neutrinos in Super-Kamiokande
and a neutrino oscillation analysis}, 
University of Tokyo, February, 1999; 
%
K. Ishihara, Doctor thesis, 
{\it Study of $\nu_\mu \to \nu_\tau$ and $\nu_mu \to \nu_{sterile}$ 
neutrino oscillations with the atmospheric neutrino data in 
Super-Kamiokande}, 
University of Tokyo, December, 1999, 
available from 
http://www-sk.icrr.u-tokyo.ac.jp/doc/sk/pub/index.html. 


\bibitem {korea}
S. B. Kim, Talk at KOSEF-JSPS Joint Seminar on New Developments
in Neutrino Physics, October 16-20, 2000, Seoul Korea,
published in Proceedings, page 182.


\bibitem {beijin}
N.~Okamura, M.~Aoki, K.~Hagiwara, Y.~Hayato, T.~Kobayashi,
T.~Nakaya and K.~Nishikawa,
{\it Prospects of very long base-line neutrino oscillation
experiments with  the JAERI-KEK high intensity proton accelerator},
hep-ph/0104220.

\bibitem {ZS88}
H.~W.~Zaglauer and K.~H.~Schwarzer,
{\it The mixing angles in matter for three generations of neutrinos
and the MSW mechanism,
Z.\ Phys.}\ C {\bf 40} (1988) 273.

\end{thebibliography}
\end{document}